\documentclass[12pt]{article}
\usepackage{axodraw}
\usepackage{epsf}

\def\theequation{\arabic{section}.\arabic{equation}}

\begin{document}
\vskip -1.5cm
\begin{flushright}
MC-TH-2002-02\\[-0.1cm] 
hep-ph/0203210\\[-0.1cm]
March 2002
\end{flushright}

\begin{center}
{\LARGE {\bf Gauge and Scheme Dependence of }}\\[0.3cm]
{\LARGE {\bf Mixing Matrix Renormalization}}\\[1.4cm]
{\large Apostolos Pilaftsis }\\[0.4cm]
{\em Department of Physics and Astronomy, University of Manchester,\\
Manchester M13 9PL, United Kingdom}
\end{center}
\vskip0.7cm  \centerline{\bf ABSTRACT} 
We revisit the issue of mixing matrix renormalization in theories that
include  Dirac or  Majorana  fermions.  We  show  how a  gauge-variant
on-shell  renormalized mixing matrix  can be  related to  a manifestly
gauge-independent  one  within a  generalized  ${\overline {\rm  MS}}$
scheme  of  renormalization.   This  scheme-dependent  relation  is  a
consequence of  the fact  that in any  scheme of  renormalization, the
gauge-dependent part of  the mixing-matrix counterterm is ultra-violet
safe  and  has  a   pure  dispersive  form.  Employing  the  unitarity
properties   of  the   theory,   we  can   successfully  utilize   the
afore-mentioned scheme-dependent relation  to preserve basic global or
local symmetries of the bare  Lagrangian through the entire process of
renormalization.  As an immediate  application of our study, we derive
the   gauge-independent  renormalization-group  equations   of  mixing
matrices in a minimal extension  of the Standard Model with isosinglet
neutrinos.

\newpage

\setcounter{equation}{0}
\section{Introduction}

One of   the  most fundamental    properties of  the  well-established
Standard  Model (SM)~\cite{GSW} is its renormalizability~\cite{Hooft}.
Renormalizability endows the  SM  with enhanced predictive  power that
emanates from the  fact that ultraviolet  (UV) divergences due to high
order quantum   effects can always   be  successfully eliminated  by a
redefinition of a finite number of independent kinematic parameters of
the theory,  such as masses and  couplings.  The predictions of the SM
have been   tested and  vindicated with   a satisfactory  accuracy  at
high-energy   colliders,  such as the   Large  Electron Positron (LEP)
collider at CERN and the Tevatron collider at  Fermilab, as well as in
low-energy experiments, e.g.\  in the  recent  E821 experiment at  BNL
where the muon anomalous magnetic moment is measured~\cite{BNL}.

In addition to masses and couplings of  the SM particles, however, the
quark-mixing  matrix, the   so-called Cabbibo-Kobayashi-Maskawa  (CKM)
matrix~\cite{CKM} $V$, needs be renormalized as well~\cite{Sirlin,DS}.
In this context,  one of the  renormalization schemes, most frequently
adopted   in  the literature,   is     the on-shell  (OS) scheme    of
renormalization~\cite{MS,AHKKM,BSH}, where  the  particle masses   are
renormalized so as  to represent the physical masses  at the  poles of
the   propagators.   It was shown   in~\cite{DS} that  the complete UV
structure of the counterterms  (CTs)  for the CKM   matrix $V$ can  be
entirely expressed  in terms  of quark wave-function renormalizations.
Within this framework, a simple  approach to renormalizing $V$ in  the
OS scheme was also presented, consistent with the unitarity properties
of the theory~\cite{GIM}.

Even  though  radiative  effects  due  to the  renormalization  of  an
off-diagonal CKM  matrix were  found to be  undetectably small  in the
SM~\cite{DS,BMP,KMS,BBS}, this  needs not be the case  for its minimal
renormalizable  extensions.   In  particular, in~\cite{KP}  the  above
formalism  of mixing-matrix renormalization  was extended  to theories
that  include  isosinglet  neutrinos  and  so admit  the  presence  of
lepton-number-violating  Majorana   masses~\cite{seesaw}.   A  minimal
realization   of  such   a  theory   is  the   SM   with  right-handed
neutrinos~\cite{SV,APzpc}.  As we will further discuss in Section~2.3,
in this minimal model the  charged and neutral current interactions of
the $W$ and  $Z$ bosons to leptons and neutrinos  are described by two
non-unitary  mixing matrices  $B$ and  $C$~\cite{APzpc}, respectively.
Most importantly,  the radiative  effects on the  light-heavy neutrino
mixing angles contained  in the $B$ and $C$  matrices were computed to
be  as   large  as  15\%~\cite{KP},  close   to  present  experimental
sensitivities.   The SM  with right-handed  neutrinos is  an appealing
scenario  which may  explain the  smallness of  the  observed neutrino
masses   and    adequately   address   the    solar   energy   deficit
problem~\cite{Concha}    through   neutrino   oscillations~\cite{MSW}.
Furthermore,  this minimal  extension of  the SM  may give  rise  to a
number  of   observable  phenomena,  such   as  lepton-flavour  and/or
lepton-number   violation   in   $\mu$,  $\tau$~\cite{IP,BKM,GC}   and
$Z$-boson  decays~\cite{LFV}, or  to possible  lepton-number violating
signals  at high-energy  colliders,  e.g.\ at  the  CERN Large  Hadron
Collider (LHC)~\cite{DGP,MMV}.

It has been noticed recently~\cite{GGM} that in the OS renormalization
prescription  presented in~\cite{DS}, the   derived  CTs for the   CKM
matrix naively depend  on the  choice  of the  gauge-fixing  parameter
$\xi$  in   the  class of   $R_\xi$  gauges.  This  fact is   not very
desirable, as physical matrix elements  will be gauge dependent  after
renormalization.   To  circumvent this problem of  $\xi$-dependence of
the   OS renormalized CKM   matrix,  several  alternative  schemes  of
renormalization    have   been    suggested  in     the    very recent
literature~\cite{GGM,KMS,BBS,YY,DK}.    As   is expected,  in all  the
proposed renormalization schemes, the UV-divergent parts of the CTs of
the CKM matrix  are identical to  those derived  in the $\overline{\rm
MS}$ scheme~\cite{DS}.   Nevertheless,  the UV-safe parts  of  the CTs
differ from approach to approach by finite dispersive constants.  Most
interestingly, one may observe  that even in the  originally suggested
OS scheme of~\cite{DS}, the gauge-dependent part of the CKM-matrix CTs
is UV finite and also has a pure  dispersive form, thus indicating the
existence of a profound  relation between gauge dependence and  scheme
dependence in mixing-matrix renormalization.

In  this paper, we revisit  the topic of mixing-matrix renormalization
of the CKM matrix $V$ and of the  $B$ and $C$ matrices. In particular,
we develop a generalized and manifestly gauge-invariant $\overline{\rm
MS}$    approach  to  mixing-matrix  renormalization.   The  developed
generalized $\overline{\rm  MS}$ approach  provides a very  convenient
framework  to address the problem  of  gauge and scheme dependences in
the  existing plethora of   differently renormalized mixing  matrices.
Moreover,  we show how our generalized  $\overline{\rm MS}$ scheme can
be successfully employed to maintain global or local symmetries of the
bare Lagrangian after renormalization.  Finally,  with the help of our
generalized   $\overline{\rm    MS}$ approach,  we   can    derive the
gauge-independent renormalization-group  (RG) equations  for    mixing
matrices.  We   explicitly demonstrate  the theoretical advantages  of
this method by calculating the one-loop RG runnings of the $B$ and $C$
matrices in the SM with isosinglet neutrinos.

The paper  is organized as follows:  after briefly reviewing the basic
formalism   of mixing  matrix renormalization   in   the OS scheme  in
Section~2.1, we present in Section~2.2 our gauge-invariant generalized
$\overline{\rm MS}$ approach to the renormalization  of the CKM matrix
$V$, and extend  it   in Section~2.3  to  the renormalization  of  the
corresponding $B$ and  $C$ mixing matrices in  the SM  with isosinglet
neutrinos.  In Section 3, we  show how our generalized  $\overline{\rm
MS}$ approach preserves additional global  and local symmetries of the
theory,  which  are  manifested   themselves as  sum   rules involving
neutrino    masses and the  $B$   and $C$ matrices.    As an immediate
application of    our considerations,  we derive    in  Section 4  the
gauge-independent  RGEs of the $B$  and $C$ mixing matrices.  Finally,
our conclusions are summarized in Section 5.

\setcounter{equation}{0}
\section{Mixing matrix renormalization}\label{sec:mix}

In this section, we will  first recall the basic analytic formulas for
the wave-function and mass CTs in the OS renormalization scheme within
the  context of general  fermionic theories,  such as  the SM  and its
natural extension  with isosinglet  neutrinos.  Then, we  will revisit
the problem of  gauge dependence of the OS-renormalized  CKM matrix in
the  SM, and  discuss its  connection  to scheme  dependence within  a
generalized   gauge-invariant  ${\overline   {\rm   MS}}$  scheme   of
renormalization.   Finally, our  discussion  will be  extended to  the
renormalization of  the mixing matrices $B$ and  $C$ that parameterize
the neutral-  and charged-current interactions in the  SM with singlet
neutrinos.

\subsection{OS renormalization scheme}

In  a theory with  a number $N_f$  of Dirac fermions, the bare kinetic
Lagrangian has the following generic form:
\begin{eqnarray}
  \label{Lkin}
{\cal L}_{\rm kin} &=&  i\, \bar{f}^0_L 
\not\! \partial f^0_L\: + \:  i\, \bar{f}^0_R \not\! \partial f^0_R\: 
-\: \bar{f}^0_L M^0 f^0_R \: -\: \bar{f}^0_R M^0 f^0_L\nonumber\\
&=&  i\, \bar{f}_L Z^{1/2\dagger}_L Z^{1/2}_L
\not\! \partial f_L\: + \:  i\, \bar{f}_R Z^{1/2\dagger}_R Z^{1/2}_R
\not\! \partial f_R\: -\: \bar{f}_L Z^{1/2\dagger}_L (M\, +\,\delta M)
Z^{1/2}_R f_R\nonumber\\
&& -\: \bar{f}_R Z^{1/2\dagger}_R (M\, +\,\delta M) Z^{1/2}_L f_L\,.
\end{eqnarray}
In the above, we have employed a matrix  notation in the space spanned
by  the  $N_f$ fermionic fields,     i.e.\ $f^T =  (f_1, f_2,   \dots,
f_{N_f})$.  As usual, we adhere the  superscript `0' to unrenormalized
quantities.   In~(\ref{Lkin}),   the   $N_f  \times  N_f$  dimensional
matrices   $Z_{L}^{1/2}$  and   $Z_{R}^{1/2}$ are  the   wave-function
renormalizations   for   the  left-   and    right-handed    fermions,
respectively.   In  addition, $M^0$, $M$ and  $\delta  M$ are diagonal
$N_f \times  N_f$ dimensional matrices  that contain the  bare masses,
the renormalized masses and their respective counter-terms (CTs).

The most general  form of an  unrenormalized $f_j \to  f_i$ transition
amplitude allowed by hermiticity~\cite{AHKKM} reads
\begin{equation}
  \label{self}
\Sigma_{ij} (\not\! p )\ =\ \not\! p P_L \Sigma^L_{ij} (p^2)\: +\:
\not\! p P_R \Sigma^R_{ij} (p^2)\: +\: P_L \Sigma^D_{ij} (p^2) \: +\:
P_R \Sigma^{D *}_{ji} (p^2)\,,
\end{equation}
supplemented by the constraints
\begin{equation}
  \label{SLR}
\Sigma^L_{ij} ( p^2 )\ =\ \Sigma^{L*}_{ji} (p^2)\,, \qquad
\Sigma^R_{ij} ( p^2 )\ =\ \Sigma^{R*}_{ji} (p^2)\, .
\end{equation}
In the  OS scheme of  renormalization, the wave-function and  mass CTs
are  given by~\cite{KP}\footnote[1]{Here,  we have  used  the symmetry
property   of  the   Lagrangian~(\ref{Lkin})  under   the  rephasings,
$Z^{1/2}_{Lij} \to e^{i\theta_i} Z^{1/2}_{Lij}$ and $Z^{1/2}_{Rij} \to
e^{i\theta_i} Z^{1/2}_{Rij}$,  in order to cast  $\delta Z^L_{ii}$ and
$\delta Z^R_{ii}$ into a symmetric  but fully equivalent form than the
one presented in~\cite{KP}.}
\begin{eqnarray}
  \label{dZL}
\delta Z^L_{ij} & = & \frac{2}{m^2_i - m^2_j}\, \Big(\, m^2_j\, 
\Sigma^L_{ij} (m^2_j )\: +\: m_i m_j \Sigma^R_{ij} (m^2_j )\:
+\: m_i \Sigma^D_{ij} (m^2_j)\: +\: m_j \Sigma^{D*}_{ji} (m^2_j)\,
\Big)\,,\\
  \label{dZR}
\delta Z^R_{ij} & = & \frac{2}{m^2_i - m^2_j}\, \Big(\, m_im_j\, 
\Sigma^L_{ij} (m^2_j )\: +\: m^2_j \Sigma^R_{ij} (m^2_j )\:
+\: m_j \Sigma^D_{ij} (m^2_j)\: +\: m_i \Sigma^{D*}_{ji} (m^2_j)\,
\Big)\,,\qquad\\
  \label{dZLdiag}
\delta Z^L_{ii} & = & -\, \Sigma^L_{ii} (m^2_i )\: +\: 
\frac{1}{2m_i}\, \Big(\, \Sigma^D_{ii} (m^2_i)\, -\, 
\Sigma^{D*}_{ii} (m^2_i)\,\Big)\nonumber\\
&& -\: m^2_i \, \Big(\, \Sigma^{L\prime}_{ii} (m^2_i )\:
+\: \Sigma^{R\prime}_{ii} (m^2_i) \, \Big)\: -\: m_i\,\Big(\,
\Sigma^{D\prime}_{ii} (m^2_i )\:
+\: \Sigma^{D*\prime}_{ii} (m^2_i) \, \Big)\,,\\
  \label{dZRdiag}
\delta Z^R_{ii} & = & -\, \Sigma^R_{ii} (m^2_i )\: -\: 
\frac{1}{2m_i}\, \Big(\, \Sigma^D_{ii} (m^2_i)\, -\, 
\Sigma^{D*}_{ii} (m^2_i)\,\Big)\nonumber\\
&& -\: m^2_i \, \Big(\, \Sigma^{L\prime}_{ii} (m^2_i )\:
+\: \Sigma^{R\prime}_{ii} (m^2_i) \, \Big)\: -\: m_i\,\Big(\,
\Sigma^{D\prime}_{ii} (m^2_i )\:
+\: \Sigma^{D*\prime}_{ii} (m^2_i) \, \Big)\,,\\
  \label{dm}
\delta m_i &=& \frac{1}{2}\, m_i\, \Big(\, \Sigma^L_{ii} (m^2_i)\: +\:
\Sigma^R_{ii} (m^2_i )\, \Big)\: +\: \frac{1}{2}\, 
\Big(\, \Sigma^D_{ii} (m^2_i)\, +\, \Sigma^{D*}_{ii} (m^2_i)\,\Big) \,, 
\end{eqnarray}
where $\Sigma^\prime (p^2) = d\Sigma  (p^2)/dp^2$ and $\delta Z^{L,R}$
are  the loop-induced  wave-function renormalizations defined  through
the relation: $Z^{1/2}_{L,R}  = 1  +  \frac{1}{2}\delta Z^{L,R}$.   We
should     bear in mind   that  only   the   dispersive  parts  of the
unrenormalized  self-energies enter the  renormalization such that the
hermiticity property of the local Lagrangian is maintained. In the SM,
it  is  $\Sigma^D_{ij}  (p^2)    =   m_i  \Sigma^S_{ij}   (p^2)$   and
$\Sigma^S_{ij}  (p^2) =  \Sigma^{S*}_{ji}  (p^2)$,  and   the formulae
(\ref{dZL})--(\ref{dm})    reduce  to  those  given  in~\cite{DS,BSH}.
However, we should stress again that these relations are very specific
to the SM and do no longer apply to extended theories.

One well-motivated  extension  of the SM is   the one in which  the SM
field content  is  augmented by  right-handed  (isosinglet) neutrinos,
thereby   admitting the     presence   of Majorana   masses   in   the
Lagrangian~\cite{seesaw,SV,APzpc}.  In this case, the fermionic fields
satisfy  the Majorana  constraints: $f^0_L   =  (f^0_R)^C$ and $f_L  =
(f_R)^C$, where the superscript  $C$ indicates charge conjugation.  As
a consequence of the Majorana constraints, we obtain the equalities:
\begin{equation}
  \label{Majconstr} 
 Z^{1/2}_L\ =\ Z^{1/2 *}_R\,,\qquad 
\Sigma^L_{ij} (p^2)\ =\ \Sigma^{R*}_{ij} (p^2)\,,\qquad 
\Sigma^M_{ij} (p^2)\ =\ \Sigma^M_{ji} (p^2)\,,
\end{equation}
where we     made  the  identification  $\Sigma^D_{ij}   (p^2)  \equiv
\Sigma^M_{ij}      (p^2)$.   Substituting    (\ref{Majconstr})    into
(\ref{dZL})--(\ref{dm})   yields  the corresponding  wave-function and
mass CTs for Majorana fields~\cite{KP}.

The issue of mixing-matrix  renormalization arises whenever one has to
deal with  the renormalization of  a non-trivial rotation  matrix that
occurs in interactions relating flavour to mass eigenstates.  To study
this problem,  we shall  adopt a perturbative  framework in  which the
classical  tree-level  Ward  identities  (WIs)  are  maintained  after
quantization.  As  such, one may consider the  Background Field Method
(BFM)~\cite{BFM,DDW}        or        the       Pinch        Technique
(PT)~\cite{PT,PP,APNPB,PPhiggs}   or    even   possible   diagrammatic
generalizations of  the latter, i.e.\ the  Generalized Pinch Technique
(GPT)~\cite{APGPT}.

\subsection{Renormalization of the CKM matrix in the SM}

As a  prototype   example,   let  us  consider    the  charged-current
interaction  in  the quark  sector of the  SM.  Specifically,  we will
revisit the renormalization  of the CKM matrix  elements~$V_{ud}$ that
enter the vertex transition $W^+(p) d(p_d) \to  u(p_u)$.  Later on, we
will   generalize       our    results     to     the   aforementioned
SU(2)$_L\otimes$U(1)$_Y$  model with  Majorana  neutrinos.  Within the
perturbative approaches mentioned above, the following tree-like WI is
satisfied~\cite{APNPB,BMP,GGM}:
\begin{eqnarray}
  \label{WI0}
p^\mu\, \Gamma_\mu^{W^+\bar{u}d,0}(p,p_u,p_d)\: 
+\: M^0_W \Gamma^{G^+\bar{u}d,0}
(p,p_u,p_d) && \nonumber\\
&&\hspace{-3cm}
=\ -\,\frac{g^0_w}{\sqrt{2}}\, 
\Big(\, V^0_{u'd}\, S^{-1,0}_{uu'} (p_u)\,P_L \: -\:  
V^0_{ud'}\, P_R\,S^{-1,0}_{d'd} (p_d)\, \Big)\,,\qquad
\end{eqnarray}
where the summation convention over  repeated quark-family indices  is
implied. In addition, in (\ref{WI0}) we have defined
\begin{eqnarray}
  \label{defs}
\Gamma_\mu^{W^+\bar{u}d,0}(p,p_u,p_d) &=&
\Gamma_{0\mu}^{W^+\bar{u}d,0}\: + \: 
\Gamma_{1\mu}^{W^+\bar{u}d}(p,p_u,p_d)\,,\\
\Gamma^{G^+\bar{u}d,0}(p,p_u,p_d) & =& 
\Gamma_{0}^{G^+\bar{u}d,0}\: + \: 
\Gamma_{1}^{G^+\bar{u}d}(p,p_u,p_d)\,,\\
S^{-1,0}_{uu'} (p_u) &=& \not\! p_u\: -\: m^0_u \: +\: 
\Sigma_{uu'} (\not\! p_u )\,, \nonumber\\
S^{-1,0}_{d'd} (p_d) &=& \not\! p_d\: -\: m^0_d \: +\: 
\Sigma_{d'd} (\not\! p_d )\,, 
\end{eqnarray}
where $\Gamma_{0\mu}^{W^+\bar{u}d,0}$ and $\Gamma_{0}^{G^+\bar{u}d,0}$
are the  bare $W^+\bar{u}d$- and  $G^+\bar{u}d$- couplings at the tree
level,    and       $\Gamma_{1\mu}^{W^+\bar{u}d}     (q,p_u,p_d)$  and
$\Gamma_1^{G^+\bar{u}d}     (q,p_u,p_d)$   are   the     corresponding
higher-order  unrenormalized        one-particle  irreducible vertices
evaluated  within e.g.\ the  PT  or the BFM.  Similar  identifications
also  apply  for the  unrenormalized  two-point  correlation functions
$S^{0}_{uu'} (p_u)$ and~$S^{0}_{d'd} (p_d)$.

Following the procedure outlined in~\cite{PPhiggs}, we require that the
same  tree-level    WI  (\ref{WI0}), which   involves   unrenormalized
quantities  only,  holds  exactly  true  after renormalization.   This
condition  can be  successfully enforced within  the gauge-independent
$\overline{\rm MS}$ scheme  of renormalization.  Nevertheless, in  any
other    favourable   scheme  of   renormalization,   the renormalized
parameters of the theory will differ  from those in the $\overline{\rm
MS}$ scheme by UV-finite  constants.  The renormalized  quantities may
be    determined in terms  of    the  unrenormalized ones through  the
relations:
\begin{eqnarray}
  \label{Rcoupl}
g^0_w &=& Z_{g_w}\, g_w\,,\nonumber\\ 
M^{02}_W &=& M^2_W\: +\: \delta M^2_W\,,\nonumber\\ 
V^0 &=& V\ +\ \delta V\,,\\
  \label{Rvert}
\Gamma_{\mu}^{W^+\bar{u}d} (p,p_u,p_d) &=& Z^{1/2}_W\, Z^{1/2\dagger}_{Luu'}\,
Z^{1/2}_{Ld'd}\, \Gamma_{\mu}^{W^+\bar{u}' d',0} (p,p_u,p_d)\,,\nonumber\\
\Gamma^{G^+\bar{u}d}(p,p_u,p_d) & =& Z^{1/2}_{G^+}\, Z^{1/2\dagger}_{uu'}\,
Z^{1/2}_{d'd}\, \Gamma^{G^+\bar{u}' d',0}(p,p_u,p_d)\,,\\
  \label{Sren}
S^{-1}_{q_iq_j} (p_q) &=& Z^{1/2\dagger}_{q_iq_k}\, S^{-1,0}_{q_kq_l}(p_q)\,
Z^{1/2}_{q_lq_j}\,,\qquad ({\rm with}\ q=u,d)\,,
\end{eqnarray}
where   $Z^{1/2}_{q_iq_j} = Z^{1/2}_{Lq_iq_j}  P_L + Z^{1/2}_{Rq_iq_j}
P_R$ and  $\delta V$ stands for  the mixing-matrix  renormalization of
the CKM matrix $V$.  In~(\ref{Rvert}), we required that $W^+\bar{u}d$-
and   $G^+\bar{u}d$-   couplings be  UV     finite after the  external
wave-function  CTs for  the $W$-boson,  the   would-be Goldstone boson
$G^+$, and the $u$- and $d$-type quarks have  been properly taken into
account.

Since our  main interest is to  compute the UV  divergent part  of the
$W^+\bar{u}d$   vertex   in the  presence of   flavour  mixing  and so
determine the UV-divergent  structure of the mixing-matrix  CT $\delta
V$, we  shall    therefore    focus   our attention   only    on   the
chirally-projected WI~(\ref{WI0})   related   to the  expression  $P_R
\Gamma_{\mu}^{W^+\bar{u}d,0} P_L$. In particular, we have 
\begin{eqnarray}
  \label{WI1}
Z^{1/2}_W\, Z^{u,1/2\dagger}_L\, P_R\, \Big( \,
p^\mu  \Gamma_{\mu}^{W^+\bar{u}d,0}\: +\: M^{02}_W\,
\Gamma^{G^+\bar{u}d,0}\, \Big)\, P_L  Z^{d,1/2}_L &&\nonumber\\
&&\hspace{-8cm}
=\ -\, Z^{1/2}_W Z_{g_w}\, \frac{g_w}{\sqrt{2}}\, Z^{u,1/2\dagger}_L\,
P_R\, \Big( \, S^{-1,0} (p_u)\, V^0\: -\: V^0\, S^{-1,0} (p_d)\, \Big)\,
P_L Z^{d,1/2}_L\, .\qquad
\end{eqnarray}
To  simplify  notation in~(\ref{WI1}),   we  have employed the  matrix
representation  for   the  quark  wave-functions  and   their  inverse
propagators,  i.e.\  $Z^{q,1/2}_L  = Z^{1/2}_{Lq_iq_j}$ and  $S^{-1,0}
(p_q)   =  S^{-1,0}_{q_i q_j}   (p_q)$, with   $q=u,d$.   Substituting
(\ref{Rvert}) and (\ref{Sren}) into (\ref{WI1}) gives
\begin{eqnarray}
  \label{WIren}
P_R\, \Big( \, p^\mu  \Gamma_{\mu}^{W^+\bar{u}d}\: +\: Z^{1/2}_W\,
Z^{-1/2}_{G^+} M^{02}_W\, \Gamma^{G^+\bar{u}d}\, \Big)\, P_L  &=&
-\, Z^{1/2}_W Z_{g_w}\, \frac{g_w}{\sqrt{2}}\nonumber\\
&&\hspace{-5.5cm}
\times P_R\, \Big( \, S^{-1} (p_u)\, Z^{u,-1/2}_L V^0 Z^{d,1/2}_L\: -\: 
Z^{u,1/2\dagger}_L V^0 Z^{d,-1/2\dagger}_L\, 
S^{-1} (p_d)\, \Big)\, P_L \, .\qquad
\end{eqnarray}
The  requirement now  that the WI   (\ref{WIren}) retains its original
form~(\ref{WI0})    where  all quantities     are replaced   by  their
renormalized ones gives rise to the following consistency conditions:
\begin{eqnarray}
  \label{SM1}
Z^{1/2}_W &=& Z^{-1}_{g_w}\,,\qquad Z_{G^+}\ =\ Z_W\, \bigg(\, 1\: +\:
\frac{\delta M^2_W}{M^2_W}\, \bigg)\,,\\
  \label{SM2}
V & =& Z^{u,-1/2}_L V^0 Z^{d,1/2}_L \ =\ Z^{u,1/2\dagger}_L V^0
Z^{d,-1/2\dagger}_L\, .
\end{eqnarray} 
The two equalities in (\ref{SM1}) are exactly  satisfied within the PT
and  BFM  frameworks~\cite{DDW,PPhiggs,GGM}.   The    double  equality
in~(\ref{SM2}) assures the   unitarity property  of  the  renormalized
mixing   matrix  $V$,   i.e.\  $V^{-1}  =   V^+$.   Most  importantly,
(\ref{SM2})  determines the analytic  structure  of the CT $\delta V$.
Employing   the  usual      decomposition    for the     wave-function
renormalizations,  i.e.\   $Z^{u,1/2}_L   =  1  +  \frac{1}{2}\,\delta
Z^{u,L}$  and $Z^{d,1/2}_L   =  1 +  \frac{1}{2}\,\delta  Z^{d,L}$, we
arrive   at   the perturbative and more    familiar  form  for $\delta
V$~\cite{DS}:\footnote[2]{Here,   we should remark  that the  analytic
results for the wave-function  CTs and $\delta  V$ obtained within the
(G)PT or  BFM are identical to  those derived  within the conventional
framework   of $R_\xi$   gauges~\cite{BBS,KMS},  with the   additional
restriction that the gauge-fixing parameters related to the photon and
the $Z$ boson are equal, i.e.~$\xi_\gamma = \xi_Z$.}
\begin{equation}
  \label{deltaV}
\delta V\ =\ \frac{1}{4}\, \Big(\,  \delta Z^{u,L}\: -\: 
\delta Z^{u,L\dagger}\, \Big)\, V\ -\  \frac{1}{4}\, V\,
\Big(\,  \delta Z^{d,L}\: -\: 
\delta Z^{d,L\dagger}\, \Big)\,,
\end{equation}
where the $u$-  and  $d$- quark wave-functions   have to satisfy   the
constraining relation~\cite{BMP,GGM}
\begin{equation}
  \label{extra}
\frac{1}{2}\, \Big(\,  \delta Z^{u,L}\: +\: 
\delta Z^{u,L\dagger}\, \Big)\, V\ = \  \frac{1}{2}\, V\,
\Big(\,  \delta Z^{d,L}\: +\: 
\delta Z^{d,L\dagger}\, \Big)\,.
\end{equation}
In the absence of flavour mixing, i.e.~for  $V =1$, this last relation
simplifies to the known one $Z^{u,L} = Z^{d,L}$~\cite{DDW,PPhiggs}.

Several important  remarks  and  observations regarding  mixing-matrix
renormalization are now in order:

\begin{itemize}

\item[(i)] The UV  poles of $\delta V$  are entirely specified by  the
wave-function   CTs of the  left-handed  $u$- and $d$-   quarks to all
orders   in perturbation  theory.  Moreover,  with   the definition of
$\delta V$ in~(\ref{deltaV}), $V$ is automatically unitary through the
order considered.

\item[(ii)] The LHS of the WI~(\ref{WIren}) is gauge-independent, when
the   chirally-projected    amplitudes  $P_R  \Gamma_\mu^{W^+\bar{u}d}
(p,p_u,p_d) P_L$  and  $P_R\Gamma^{G^+\bar{u}d}  (p,p_u,p_d) P_L$  are
evaluated by setting the  external   particles on their mass   shells.
Consequently,  the RHS  of~(\ref{WIren}) must  be gauge-independent as
well.    This  can  only happen,  if   $V$ and  hence   $\delta V$ are
gauge-independent~\cite{BMP,GGM}.  For    example,   unlike in     the
${\overline{\rm    MS}}$   scheme~\cite{DS},    $\delta         V$  is
$\xi$-dependent~\cite{KMS,BBS,YY,DK}      in    the    OS  scheme   of
renormalization.  As we will see  below in~(iii), however, because the
gauge-dependent  part  of  $\delta V$  is  UV  finite  and has  a pure
dispersive  form, the $\xi$-dependent  terms of an OS renormalized CKM
matrix $V$ can always be related to finite gauge-independent constants
in a generalized and  manifestly gauge-invariant ${\overline{\rm MS}}$
scheme of renormalization.

\item[(iii)]  There   exists     an  underlying  symmetry   in     the
renormalization   of   $V^0$, reflecting the   presence  of  a general
intrinsic  freedom   in     redefining mixing   matrices  at    higher
orders.\footnote[3]{The  existence of such  a scheme dependence in the
renormalization   of $V^0$ was first  pointed  out in~\cite{KP}.}  The
presence   of   this    higher-order scheme   arbitrariness     in the
renormalization of  $V^0$ may be described  as follows.  We  know that
the CKM matrix is the product of  two unitary matrices $U^{u,0}_L$ and
$U^{d,0}_L$ relating  the weak to mass  eigenstates of the left-handed
$u$-  and $d$-type quark fields,  respectively, i.e.\ $V^0 = U^{u,0}_L
U^{d,0\dagger}_L$. If we now perform the following perturbative shifts
in the left-handed quark wave functions and their mixing-matrix CTs:
\begin{eqnarray}
  \label{ZudL}
\delta Z^{u,L} &\to & \delta Z^{u,L}\: +\: c^u\,,\qquad \quad\ \,
\delta Z^{d,L} \ \to \ \delta Z^{d,L}\: +\: c^d\,,\\
  \label{UudL}
\delta U^u_L & \to & \delta U^u_L \: -\: \frac{1}{2}\, c^u\,
U^u_L\,,\qquad
\delta U^{d\dagger}_L \ \to \ \delta U^{d\dagger}_L \: +\: \frac{1}{2}\,
U^{d\dagger}_L\, c^d\,,\qquad
\end{eqnarray}
where $c^u$ and $c^d$  are anti-hermitian, gauge-independent UV-finite
constant matrices, i.e.\  $c^{u,d} = -  c^{u,d\dagger}$ and $c^{u,d} =
{\cal  O}(\delta Z^{u,L},\delta Z^{d,L})$,  then none of the important
key  equalities~(\ref{SM2}),   (\ref{deltaV})  and (\ref{extra})  will
change through  the   order  considered. On the  basis   of  the above
unitarity symmetry,     we  may  generally    define  the   manifestly
gauge-independent CKM matrix CT:
\begin{equation}
  \label{dVgen}
\delta V\ =\ \delta V^{\overline{\rm MS}}\: -\: \frac{1}{2}\, c^u\,
V\: +\:  \frac{1}{2}\, V\, c^d\, , 
\end{equation}
where $\delta V^{\overline{\rm MS}}$  is the  corresponding CT in  the
gauge-invariant    ${\overline{\rm  MS}}$     scheme         evaluated
in~(\ref{deltaV}).  Evidently, $\delta V$  does not give rise to gauge
dependences  in  the  computation  of physical transition  amplitudes,
e.g.\ in the top-decay amplitude $t \to W+ b$.  However, the UV-finite
constants $c^u$ and    $c^d$ introduce a  scheme  dependence  into the
renormalization of the CKM matrix $V^0$. Thus, all the various schemes
of   renormalization  proposed   in the    literature~\cite{DS,KP,GGM,
KMS,BBS,YY,DK},   including   the  OS  scheme~\cite{DS,KP},   may   be
represented  by   the gauge-invariant  expression~(\ref{dVgen})   with
appropriate choices for the UV-finite matrices $c^u$ and $c^d$.

Finally, it  is interesting to notice  that the number  of independent
parameters contained in the anti-hermitian matrices $c^u$ and $c^d$ is
exactly equal to the number  of group parameters, the so-called mixing
angles, that generate the   unitary rotations in the left-handed  $u$-
and $d$- quark  flavour space.  To  elucidate the above point, let  us
consider the renormalized  unitary matrix in the left-handed $u$-quark
flavour space, $U^u_L$.  In particular, $U^u_L$ can be written as
\begin{equation}
  \label{UuLgroup}
U^u_L (\theta^a_u)\ =\ \exp\, \Big( i\,\theta^a_u\, T^a\,\Big)\,,
\end{equation}
where  $\theta^a_u$ are   the  group  parameters  and  $T^a$ are   the
associate generators of the U$(n)$ flavour group, satisfying the usual
Lie-algebra relations: $[T^a,T^b] =    i f^{abc}\,T^c$, with   $T^a  =
T^{a\dagger}$, $T^0 = {\bf 1}_n$  and $f^{0ab} =0$.   If we now  shift
$\theta^a_u$ by a finite higher-order amount $\delta \theta^a_u$, then
the unitary matrix~(\ref{UuLgroup}) exhibits the following variation:
\begin{eqnarray}
  \label{UuLdiff}
U^u_L (\theta^a_u + \delta \theta^a_u )
&=& \bigg[\, {\bf 1}_n\: +\: i\, \bigg(\, \delta^{bc}\: -\:
\frac{1}{2}\, f^{bcd}\theta^d_u\nonumber\\
&-&\!\!\! 
\frac{1}{6}\, f^{bdx}f^{cex} \theta^d_u \theta^e_u\ +\ \cdots \bigg)\, 
\delta\theta^b_u T^c\, \bigg]\: 
U^u_L (\theta^a_u)\ +\ {\cal O}\Big(\delta\theta^{a\,2}_u\Big)\, .
\end{eqnarray}
In   writing  the  RHS  of  (\ref{UuLdiff}),  we    have employed  the
Baker-Hausdorff  formula   for  infinitesimal  non-Abelian  rotations.
Comparing  (\ref{UuLdiff}) with (\ref{UudL}), we immediately recognize
that the anti-hermitian matrix $c^u$ is given by
\begin{equation}
  \label{cu}
c^u \ =\ -\,2i\, \bigg(\, \delta^{ab}\: -\:
\frac{1}{2}\, f^{abc}\theta^c_u\: -\:  
\frac{1}{6}\, f^{acx}f^{bdx} \theta^c_u \theta^d_u\ +\ \cdots \bigg)\, 
\delta\theta^a_u T^b\,,
\end{equation}
where  the   $\delta\theta^a_u$'s  parameterize  the  scheme-dependent
shifts   in  the   mixing  angles   $\theta^a_u$.   Notice   that  the
anti-hermitian  matrix   $c^u$  in~(\ref{cu}),  determined   by  means
of~(\ref{UuLdiff}), preserves  the unitarity properties  of $U^u_L$ by
construction.  Analogous  determinations may  also be found  for $c^d$
and hence for the scheme-dependent part of $V$ in~(\ref{dVgen}).

\end{itemize}

\subsection{Mixing renormalization in the SM with singlet neutrinos}

We will now discuss the problem of mixing-matrix renormalization in an
SU$(2)\otimes$U$(1)_Y$  theory  with   a  number  $N_G$  of  fermionic
doublets and  a number  $N_R$ of right-handed  (isosinglet) neutrinos.
The interaction Lagrangians of this model which describe the couplings
of  the $W^\pm$,  $Z$  and Higgs  ($H$)  bosons to  the $N_G$  charged
leptons, $l_i$, and $(N_G + N_R)$ Majorana neutrinos, $n_i$, are given
by
\begin{eqnarray}
  \label{LW}
{\cal L}_W &=& -\, \frac{g_w}{\sqrt{2}}\, W^-_\mu\, \bar{l}_i\, B_{ij}
\gamma^\mu P_L\, n_j\ +\ {\rm h.c.},\\
  \label{LZ}
{\cal L}_Z &=& -\, \frac{g_w}{4 c_w}\, Z_\mu\, \bar{n}_i \, \gamma^\mu\,
\Big(\,C_{ij}\, P_L\: -\: C^*_{ij}\,P_R\,\Big)\, n_j\, \\
  \label{LH}
{\cal L}_H &=& -\,\frac{g_w}{4M_W}\, H\, \bar{n}_i \, 
\Big[\, \Big(\, m_i C_{ij}\: +\: m_j C^*_{ij}\,\Big) P_L\: +\: 
\Big(\, m_i C^*_{ij}\: +\: m_j C_{ij}\, \Big)\, P_R\,\Big]\, n_j\, .
\end{eqnarray}
Here,    we    follow   the   conventions    of~\cite{APzpc,IP}.    In
(\ref{LW})--(\ref{LH}), $B$  and  $C$  are  $N_G\times (N_G+N_R)$  and
$(N_G+N_R)\times (N_G+N_R)$-dimensional mixing matrices, defined as
\begin{eqnarray}
 \label{Blj}
B_{lj} &=& \sum\limits_{k=1}^{N_G}\,  V^l_{lk}\, U^{n *}_{kj}\,,\\
 \label{Cij}
C_{ij} &=& \sum\limits_{k=1}^{N_G}\,  U^{n}_{ki}\, U^{n *}_{kj}\,.
\end{eqnarray}
In~(\ref{Blj}), the $N_G\times N_G$ unitary matrix $V^l$ occurs in the
diagonalization of the charged-lepton mass matrix and relates the weak
to the mass   eigenstates     of the left-handed   charged    leptons.
Correspondingly,  the   $(N_G+N_R)\times   (N_G+N_R)$ unitary   matrix
$U^n$ in~(\ref{Cij}) diagonalizes the symmetric neutrino-mass matrix
\begin{equation}
  \label{Mnu}
M^n \ =\ \left(\! \begin{array}{cc}
0 & m_D \\ m^T_D & m_M \end{array} \!\right)\, ,
\end{equation}
through the unitary transformation
\begin{equation}
  \label{Umass}
U^{n T}\, M^n\, U^n\ =\ \widehat{M}^n\ =\ 
{\rm diag}\, \Big(\,m_1,\ m_2,\ \dots,\ 
m_{N_G+N_R}\,\Big)\, .
\end{equation}
At  this point, we  should  recall again~\cite{APzpc} that  the mixing
matrices  $B$ and $C$  obey a number of  basic identities which ensure
the renormalizability of the theory:
\begin{eqnarray}
  \label{Id1}
\sum_{k=1}^{N_G +N_R}\, B_{lk}\, B^*_{l'k} &=& \delta_{ll'}\,,\\
  \label{Id2}
\sum_{k=1}^{N_G +N_R}\, C_{ik}\, C_{kj} &=& C_{ij}\,,\\
  \label{Id3}
\sum_{k=1}^{N_G +N_R}\, B_{lk}\, C_{ki} &=& B_{li}\,,\\
  \label{Id4}
\sum_{l=1}^{N_G}\, B^*_{li}\, B_{lj} &=& C_{ij}\,,\\
  \label{Idl1}
\sum_{k=1}^{N_G + N_R}\, m_k\, B_{lk}\, B_{l'k} &=& 0\,,\\
  \label{Idl2}
\sum_{k=1}^{N_G + N_R}\, m_k\, B_{lk}\, C_{ik} &=& 0\,,\\
  \label{Idl3}
\sum_{k=1}^{N_G + N_R}\, m_k\, C_{ik}\, C_{jk} &=& 0\,.
\end{eqnarray}
The last three relations~(\ref{Idl1})--(\ref{Idl3}) are manifestations
of the  presence of  lepton-number  violation in the  neutrino sector.
Instead, if theory conserves lepton number, these three identities are
not necessary. In this case,  Majorana neutrinos are either degenerate
in  pairs forming massive Dirac neutrinos  or unpaired  giving rise to
massless  Majorana-Weyl  two-component spinors.   As a  consequence of
lepton-number conservation, the mixing  matrix elements $C^*_{ij}$ are
absent from the $Zn_in_j$- and $H n_in_j$- couplings in~(\ref{LZ}) and
(\ref{LH}),     so  the  $Zn_in_j$-coupling   becomes  purely  chiral,
proportional to the operator $\bar{n}_i \gamma_\mu P_L n_j$.

The renormalization of the mixing matrices $B$  and $C$ can be carried
out in a  way very similar to  the SM case.  Following analogous steps
for the one-particle  irreducible vertex functions $\Gamma^{W^-\bar{l}
n_j}$ and $\Gamma^{Z\bar{n}_i n_j}$ in the BFM or PT, we find
\begin{eqnarray}
  \label{Bren}
B & =& Z^{l, -1/2}_L\, B^0\, Z^{n,1/2}_L\ =\ Z^{l, 1/2\dagger}_L\, B^0\,
Z^{n,-1/2\dagger}_L\,, \\
  \label{Cren}
C & =& Z^{n, -1/2}_L\, C^0\, Z^{n,1/2}_L\ =\ Z^{n, 1/2\dagger}_L\, C^0\,
Z^{n,-1/2\dagger}_L\,, 
\end{eqnarray}
where   $Z^{l,  1/2}_L$      and $Z^{n,1/2}_L$   are     wave-function
renormalization  matrices for  the  left-handed   charged leptons  and
Majorana   neutrinos,   respectively.    Equations~(\ref{Bren})    and
(\ref{Cren}) lead perturbatively to the mixing-matrix CTs~\cite{KP}
\begin{eqnarray}
  \label{dB}
\delta B &=& \frac{1}{4}\, \Big(\, \delta Z^{l,L}\: -\:
\delta Z^{l,L\dagger}\,\Big)\, B\ -\ \frac{1}{4}\, B\,
\Big(\, \delta Z^{n,L}\: -\: \delta Z^{n,L\dagger}\,\Big)\,,\\
  \label{dC}
\delta C &=& \frac{1}{4}\, \Big(\, \delta Z^{n,L}\: -\:
\delta Z^{n,L\dagger}\,\Big)\, C\ -\ \frac{1}{4}\, C\,
\Big(\, \delta Z^{n,L}\: -\: \delta Z^{n,L\dagger}\,\Big)\,.
\end{eqnarray}
In addition, the following constraining relations are satisfied:
\begin{eqnarray}
  \label{Bcon}
\frac{1}{2}\, \Big(\, \delta Z^{l,L}\: +\:
\delta Z^{l,L\dagger}\,\Big)\, B &=& \frac{1}{2}\, B\,
\Big(\, \delta Z^{n,L}\: +\: \delta Z^{n,L\dagger}\,\Big)\,,\\
  \label{Ccon}
\frac{1}{2}\, \Big(\, \delta Z^{n,L}\: +\:
\delta Z^{n,L\dagger}\,\Big)\, C &=&  \frac{1}{2}\, C\,
\Big(\, \delta Z^{n,L}\: +\: \delta Z^{n,L\dagger}\,\Big)\,.
\end{eqnarray}
It is  important to observe  that the renormalized mixing matrices $B$
and  $C$   given  in~(\ref{Bren}) and~(\ref{Cren})   as   well  as the
perturbative  definitions of  the mixing-matrix   CTs  $\delta B$  and
$\delta C$  by means  of  (\ref{dB}) and  (\ref{dC}) fully satisfy the
identities~(\ref{Id1})--(\ref{Id4}).  However, the  compatibility   of
$\delta     B$     and    $\delta     C$      with    the    remaining
identities~(\ref{Idl1})--(\ref{Idl3}) proves    more  subtle and  will
be discussed in Section~\ref{sec:sum}.

In the ${\overline {\rm MS}}$ scheme, the mixing-matrix CTs $\delta B$
and $\delta C$ become  gauge independent, only after the corresponding
gauge-dependent     part     of     the     tadpole     graphs     are
included~\cite{PT,PP,DDW,YY}. To further  illuminate our procedure, we
calculate  the Higgs-boson  tadpole  graph $\Gamma^H$  induced by  the
$W^+$  boson, and  the associate  would-be Goldstone  boson  $G^+$ and
ghost fields~$c^+,\bar{c}^+$, i.e.
\begin{eqnarray}
  \label{TadH}
\Gamma^{H}_{(W)} (0) &=& \frac{g_w}{32\pi^2}\,
\frac{M^2_H}{M_W}\, \Big[\, \xi_W M^2_W \, 
\Big(\, 1\: +\: B_0 (0,\xi_W M^2_W,\xi_W M^2_W)\, \Big)\,\Big]\nonumber\\
&&+\, \frac{g_w}{16\pi^2}\,(D-1)\,M_W\, \Big[\, M^2_W \, 
\Big(\, 1\: +\: B_0 (0,M^2_W,M^2_W)\, \Big)\,\Big]\, ,
\end{eqnarray}
where  the  one-loop  function $B_0(p^2,m^2_1,m^2_2)$   is  defined in
Appendix    A.  {}From~(\ref{TadH}) it is easy    to see that only the
$M^2_H$-dependent part of the tadpole depends on $\xi_W$ and should be
included   in  the  scalar   part   of   the  self-energy  transitions
$\Sigma^D_{ij}$    (cf.~(\ref{self})).      More       precisely,  the
$M^2_H$-dependent part of  the   tadpole graph effectively induces   a
gauge-dependent shift to the $H$-boson VEV $v$:
\begin{equation}
  \label{dVEVw}
\bigg(\frac{\delta v}{v}\bigg)^{\xi_W}\ =\ \frac{g_w}{2M_W}\, 
\frac{(\Gamma^H)^{\xi_W}}{M^2_H}\ =\ \frac{\alpha_w}{16\pi}\,  \xi_W \, 
\Big(\, 1\: +\: B_0 (0,\xi_W M^2_W,\xi_W M^2_W)\, \Big)\,.
\end{equation}
Similarly,  the  $Z$-boson loop  causes an  analogous  gauge-dependent
shift to $v$, i.e.
\begin{equation}
  \label{dVEVz}
\bigg(\frac{\delta v}{v}\bigg)^{\xi_Z}\ =\ \frac{g_w}{2M_W}\, 
\frac{(\Gamma^H)^{\xi_Z}}{M^2_H}\ =\ \frac{\alpha_w}{32\pi\,c^2_w}\,  
\xi_Z \, \Big(\, 1\: +\: B_0 (0,\xi_Z M^2_Z,\xi_W M^2_Z)\, \Big)\,.
\end{equation}
Then, the $\xi$-dependent  VEV shifts~(\ref{dVEVw}) and  (\ref{dVEVz})
contribute  the  following  term     to   the scalar  part    of   the
Majorana-neutrino self-energy transitions:
\begin{equation}
  \label{Stad} 
\Sigma^{M,{\rm tad}}_{ij}\,P_L\ =\ -\, 
\bigg(\frac{\delta v}{v}\bigg)^{\xi_{W,Z}}\,
\Big(\, m_i C_{ij}\: +\: m_j C^*_{ij}\, \Big)\, P_L\, .
\end{equation}
Note that if  neutrinos are Dirac particles, one has  just to drop the
term  proportional to $C^*_{ij}$  on the  RHS of~(\ref{Stad}).   As we
will  see   more  explicitly  in   the  next  sections,   the  tadpole
contribution (\ref{Stad})  plays an instrumental r\^ole  to render the
mixing-matrix CTs $\delta B$ and $\delta C$ gauge independent.

\setcounter{equation}{0}
\section{Neutrino mass-mixing sum rules}\label{sec:sum}

As  was   already mentioned  in   Section~\ref{sec:mix}, the  neutrino
mass-mixing sum rules~(\ref{Idl1})--(\ref{Idl3}) are very essential to
ensure  the  renormalizability  of the   theory.  These  sum rules are
obtained  by projecting out the  zero texture in the Majorana-neutrino
mass matrix~(\ref{Mnu}) as follows:
\begin{equation}
 \label{mixsum}
\sum\limits_{k=1}^{N_G+N_R}\, m_k\, U^{n}_{lk}\, U^{n}_{l'k}\ =\
\big(\,U^n\, \widehat{M}^n U^{n T}\,\big)_{ll'}\ =\
M^{n *}_{ll'}\ =\ 0\,, 
\end{equation}
for $l,l' =1,2,\dots,N_G$.  The zero texture is protected by the gauge
symmetry   of  the  theory,   since  the   contributing  5-dimensional
gauge-invariant operator  $\bar{L}^T \Phi^T  \Phi L^C$ is  absent from
the  local renormalizable  Lagrangian, where  $L$ and  $\Phi$  are the
lepton and Higgs doublets, respectively.  This operator is radiatively
generated at the  one-~\cite{CW,APzpc} and two-~\cite{EMa} loop levels
and is UV finite.  The neutrino mass-mixing sum rule~(\ref{mixsum}) is
no  longer valid,  if  the theory  is  extended by  one Higgs  triplet
$\Delta_L$, since the afore-mentioned lepton-number-violating operator
can  now  appear  in  the  tree  level  Lagrangian  through  the  term
$\bar{L}^T \Delta_L L^C$.

In the following, we will show that  renormalization of $U^n$ does not
spoil the basic identity~(\ref{mixsum}) in  the ${\overline {\rm MS}}$
scheme.      Within the    scheme   of   renormalization outlined   in
Section~\ref{sec:mix}, the CT matrix $\delta U^n$ of $U^n$ is given by
\begin{equation}
  \label{dU}
\delta U^n\ =\ \frac{1}{4}\, U^n \,\Big(\,\delta Z^{n,LT}\: -\:
\delta Z^{n,L *}\,\Big)\, .
\end{equation}
Observe  that  (\ref{dU}) may  also be derived   by setting $V^l  = 1$
in~(\ref{dB}). In order that the bare and renormalized mixing matrices
$U^{n,0}$ and $U^n$ obey~(\ref{mixsum}), one has to show that
\begin{equation}
  \label{CTsum}  
\big(\,\delta  U^n\, \widehat{M}^n\,  U^{n T}\:
+\:  U^n\, \widehat{M}^n\, \delta  U^{n  T}\: +\: U^n\, \delta
\widehat{M}^n\, U^{n T}\,\big)_{ll'}\ =\ 0\, ,
\end{equation}
namely the corresponding mixing  and mass CTs obey also~(\ref{mixsum})
up to  higher orders  of  perturbation proportional to  $(\delta U^n
)^2$.

Before offering  a proof of~(\ref{CTsum})  for the most  general case,
let  us first  gain some  insight from  considering  an one-generation
model with one right-handed neutrino  only.  The mass spectrum of this
simple   model,   which  essentially   resembles   the  known   seesaw
scenario~\cite{seesaw},  consists of two  Majorana neutrinos:  a light
neutrino $\nu$ observed in  experiment and a yet-undetected superheavy
one~$N$.  Most interestingly,  in this model the elements  of the bare
mixing matrix $U^{n,0}$  are entirely determined by the  two bare mass
eigenvalues, $m^0_\nu$ and $m^0_N$, of $M^{n,0}$ in~(\ref{Mnu}):
\begin{equation}
  \label{U1gen}
U^0_{\nu\nu}\ =\ \sqrt{\frac{m^0_N}{m^0_\nu + m^0_N}}\, ,\quad
U^0_{\nu N}\ =\ i\, \sqrt{\frac{m^0_\nu}{m^0_N}}\, U^0_{\nu\nu}\, , \quad
U^0_{N\nu} \ =\ U^0_{\nu N}\,,\quad U^0_{NN}\ =\ U^0_{\nu\nu}\, ,
\end{equation}
where we have dropped the superscript  `$n$' from $U^n$ and have chosen
the  phase  convention   in which   the  elements $U^0_{\nu\nu}$   and
$U^0_{NN}$   are positive. In the  ${\overline  {\rm MS}}$ scheme, the
mass, wave-function and mixing CTs are found to be
\begin{eqnarray}
  \label{MSdm}
\delta m_\nu &=& -\, \frac{\alpha_w}{16\pi}\, \frac{m_\nu m_N}{m_\nu
+m_N}\, \bigg(\, \frac{3m^2_l}{M^2_W}\: -\: \frac{m^2_\nu}{M^2_W}\:
-\: \frac{m_\nu m_N}{M^2_W}\, \bigg)\, C_{\rm UV}\,,\nonumber\\ 
\delta m_N &=& -\, \frac{\alpha_w}{16\pi}\, \frac{m_\nu m_N}{m_\nu
+m_N}\, \bigg(\, \frac{3m^2_l}{M^2_W}\: -\: \frac{m^2_N}{M^2_W}\:
-\: \frac{m_\nu m_N}{M^2_W}\, \bigg)\, C_{\rm UV}\,,\\
  \label{MSdZ}
\delta Z^{L}_{\nu\nu} &=& -\, \frac{\alpha_w}{16\pi}\, C_{\nu\nu}\, \bigg(\,
2\xi_W\: +\: \frac{\xi_Z}{c^2_w}\: +\: \frac{m^2_l}{M^2_W}\: +\:
\frac{m^2_\nu}{M^2_W}\: +\: \frac{m_\nu m_N}{M^2_W}\, \bigg)\,  C_{\rm
UV}\,,\nonumber\\
\delta Z^{L}_{\nu N} &=& -\, \frac{\alpha_w}{8\pi}\, C_{\nu N}\,
\bigg(\,\xi_W\: +\: \frac{\xi_Z}{2c^2_w}\: -\:
\frac{m^2_l}{M^2_W}\: +\: \frac{3m^2_l}{M^2_W}\, 
\frac{m_\nu}{m_\nu + m_N}\, \bigg)\,  C_{\rm UV}\,,\nonumber\\
\delta Z^{L}_{N\nu} &=& -\, \frac{\alpha_w}{8\pi}\, C_{N\nu}\,
\bigg(\,\xi_W\: +\: \frac{\xi_Z}{2c^2_w}\: -\: \frac{m^2_l}{M^2_W}\: +\:
\frac{3m^2_l}{M^2_W}\, \frac{m_N}{m_\nu + m_N}\, \bigg)\,  
C_{\rm UV}\,,\nonumber\\
\delta Z^{L}_{NN} &=& -\, \frac{\alpha_w}{16\pi}\, C_{NN}\, \bigg(\,
2\xi_W\: +\: \frac{\xi_Z}{c^2_w}\: +\: \frac{m^2_l}{M^2_W}\: +\:
\frac{m^2_N}{M^2_W}\: +\: \frac{m_\nu m_N}{M^2_W}\, \bigg)\,  C_{\rm
UV}\,,\\
  \label{MSdU}
\delta U_{\nu\nu} &=& \delta U_{NN}\ =\ U_{\nu\nu}\,\frac{3\alpha_w}{32\pi}\, 
\frac{m_\nu\, (m_N - m_\nu)}{(m_\nu + m_N)^2}\, \frac{m^2_l}{M^2_W}\,  
C_{\rm UV}\,,\nonumber\\
\delta U_{\nu N} &=& \delta U_{N\nu}\ =\ 
-\,U_{\nu N}\,\frac{3\alpha_w}{32\pi}\, 
\frac{m_N\, (m_N - m_\nu)}{ (m_\nu + m_N)^2}\, \frac{m^2_l}{M^2_W}\,  
C_{\rm UV}\,.
\end{eqnarray}
In   the  above,   $C_{ij}  =   U_{\nu  i}   U^*_{\nu   j}$  according
to~(\ref{Cij})  (with $i,j =  \nu,\ N$),  $m_l$ is  the charged-lepton
mass, and  $C_{\rm UV}$ is an  UV constant defined in  Appendix A.  As
was  discussed in  Section~\ref{sec:mix}, we  find that  the  mass CTs
$\delta  m_\nu$ and  $\delta  m_N$,  and the  CT  matrix $\delta  U^n$
computed  by~(\ref{dU}) are gauge-independent  only after  the tadpole
contributions  are included.   Moreover, these  CTs satisfy  the basic
identity~(\ref{CTsum}), i.e.
\begin{equation}
  \label{CTsum1}
\delta m_\nu\, U^2_{\nu\nu} \: +\: \delta m_N\, U^2_{\nu N} \: +\: 
2m_\nu\, U_{\nu\nu}\,\delta U_{\nu\nu}\: +\: 2m_N\,
U_{\nu N}\,\delta U_{\nu N}\ =\ 0 \, .
\end{equation}

In  the above simple Majorana-neutrino model,  it is still possible to
follow an alternative approach.  Specifically, the same  results would
have  been obtained if we had  considered the elements of $U^{n,0}$ as
functions of $m^0_\nu$ and $m^0_N$, i.e.\ $U^{n,0} = U^{n,0} (m^0_\nu,
m^0_N)$.    In   this case,  the mixing-matrix   CT   $\delta  U^n$ is
calculated as~\cite{DK}
\begin{equation}
  \label{dUDK}
\delta U^n\ =\ \delta m_\nu \,\frac{\partial U^n (m_\nu, m_N)}{\partial
m_\nu}\: +\: \delta m_N \,\frac{\partial U^n (m_\nu, m_N)}{\partial
m_N}\ ,
\end{equation}
and the basic relation~(\ref{CTsum1}) of  the CTs will be satisfied by
construction,  even  within  the  OS scheme  of  renormalization.   In
addition, $\delta U^n$ defined in  terms of $\delta m_\nu$ and $\delta
m_N$   is    gauge-independent~\cite{DK}.    Instead,   if    we   had
employed~(\ref{dU})  to compute  $\delta U^n$  in the  OS  scheme, the
resulting expression would have  naively been gauge-dependent and have
violated    the    CT    relation~(\ref{CTsum1})    by    UV    finite
terms.\footnote[4]{This    last    result    was   earlier    observed
in~\cite{KP}.}  Nevertheless,  even if the  mass CTs are  evaluated in
the  OS  scheme,  we  can  always  restore the  validity  of  the  sum
rule~(\ref{mixsum}),       along      with       the      constraining
relation~(\ref{CTsum}),    by   redefining    $\delta   U^n$    in   a
gauge-invariant  manner.   To  be  specific,  exactly  as  we  did  in
Section~\ref{sec:mix},  we  add   a  gauge-independent  and  UV-finite
anti-hermitian  matrix $c^n$  to  the ${\overline  {\rm  MS}}$ CTs  of
$U^n$:
\begin{equation}
  \label{Ucn}
\delta U^n\ =\ \delta U^{n,\overline {\rm MS}}\ +\ \frac{1}{2}\,
c^n\, U^n\,,
\end{equation}
with $c^n = - c^{n\dagger}$.   In agreement with our phase conventions
for the matrix elements of $U^n$ in~(\ref{U1gen}), it is sufficient to
assume that the matrix $c^n$ takes on the form
\begin{equation}
  \label{cn}
c^n \ =\ \left( \!
\begin{array}{cc} 0 & c \\ -c & 0 \end{array} \! \right)\,,
\end{equation}
where $c$ is a real constant.  Then, the parameter $c$ can be uniquely
determined by requiring that the constraining relation~(\ref{CTsum1}),
with the mass CTs  $\delta m_\nu$ and $\delta m_N$  computed in the OS
scheme, holds exactly true.  In this way, we  were able to verify that
the  so-derived mixing-matrix  CTs  $\delta   U_{\nu\nu}$ and  $\delta
U_{\nu N}$ are identical with those obtained by virtue of~(\ref{dUDK}).

For the more realistic  case, for which the SM  contains more than one
right-handed  neutrino, the unitary  matrix  $U^n$ cannot be  entirely
expressed  in terms of neutrino masses,   and the alternative approach
based on~(\ref{dUDK}) turns out to be not very practical. Instead, one
may   utilize the more  general  approach   described above,  in which
anti-hermitian constants $c^n$  are added to the ${\overline{\rm MS}}$
CT  $\delta  U^n$.    Within this    generalized   framework of    the
${\overline{\rm MS}}$ scheme, the constraining relations~(\ref{CTsum})
reduce  the  number  of independent  constants   $c^n$.  The remaining
freedom should  be fixed by  comparing the theoretical predictions for
observables involving the undetermined  matrix elements of $U^n$  with
experiment.  Here, we  should   stress again that   the anti-hermitian
constants $c^n$      are  only    required   if  the     neutrino-mass
renormalizations $\delta  m_i$  are computed  by~(\ref{dm}) in  the OS
scheme.  

Unlike in the OS  scheme, in the  ${\overline{\rm MS}}$ scheme one has
to pay the price  that the ${\overline{\rm MS}}$-renormalized neutrino
masses are  not the physical ones,   namely the poles  of the neutrino
propagators. However, as we will now show, all the basic symmetries of
the  theory,    including    the   one    reflected    in  the     sum
rule~(\ref{mixsum}), are preserved  and the addition of anti-hermitian
constants $c^n$ is no longer needed.  In particular, we will provide a
general    proof  of     the     validity    of   the     constraining
relation~(\ref{CTsum}) governing the neutrino- mass  and mixing CTs in
the  ${\overline{\rm    MS}}$   scheme.   To  this    end,  we   first
substitute~(\ref{dU}) into~(\ref{CTsum})
\begin{eqnarray}
 \label{proof1}
&&\hspace{-2cm} 
\delta  U^n\, \widehat{M}^n\,  U^{n T}\:
+\:  U^n\, \widehat{M}^n\, \delta  U^{n  T}\: +\: U^n\, \delta
\widehat{M}^n\, U^{n T}\,\Big|_{ll'}\nonumber\\
&=&  U^n\, \bigg[\, 
\frac{1}{4}\,\big(\delta Z^{n,L T}\, -\, 
\delta Z^{n,L*}\big)\, \widehat{M}^n\: +\:  
\frac{1}{4}\, \widehat{M}^n\, \big(\delta Z^{n,L}\, -\, 
\delta Z^{n,L\dagger}\big)\: +\: \delta
\widehat{M}^n\,\bigg] U^{n T}\,\Big|_{ll'}\nonumber\\
&=& U^n\, \bigg(\, 
\frac{1}{2}\,\delta Z^{n,L T}\,\widehat{M}^n\: +\:  
\frac{1}{2}\, \widehat{M}^n\, \delta Z^{n,L}\: +\: \delta
\widehat{M}^n\,\bigg) U^{n T}\,\Big|_{ll'}\nonumber\\
&&-\, \frac{1}{4}\, 
U^n\, \Big[\, \big(\delta Z^{n,L T}\, +\, 
\delta Z^{n,L*}\big)\, \widehat{M}^n\: +\:  
\widehat{M}^n\, \big(\delta Z^{n,L}\, +\, 
\delta Z^{n,L\dagger}\big)\,\Big] U^{n T}\,\Big|_{ll'}\nonumber\\
&=& U^n\, \Sigma^{M *,{\rm UV}}\, U^{n T}\,\Big|_{ll'}\ +\
\frac{1}{2}\, U^n\,\Big( \Sigma^{L T,{\rm UV}}\,\widehat{M}^n\: +\:
\widehat{M}^n\, \Sigma^{L,{\rm UV}}\,\Big)\,U^{n T}\,\Big|_{ll'}\, ,
\end{eqnarray}
where the superscript UV  on $\Sigma^M$ and $\Sigma^L$ indicates their
UV divergent parts, and $l,l' =  1,2,\dots,N_G$.  In deriving the last
step of~(\ref{proof1}), we have used the relations
\begin{eqnarray}
  \label{SMSL}
\Sigma^{M,{\rm UV}} &=& \delta \widehat{M}^n \ +\ 
\frac{1}{2}\,  \widehat{M}^n\, \delta Z^{n,L*}\ +\ 
\frac{1}{2}\,  \delta Z^{n,L\dagger}\,\widehat{M}^n\, ,\\
\Sigma^{L,{\rm UV}} &=& -\, \frac{1}{2}\, \Big(\,\delta Z^{n,L}\ +\ 
\delta Z^{n,L\dagger}\,\Big)\, .
\end{eqnarray}
These relations  may    be straightforwardly obtained    with  the aid
of~(\ref{dZL}) and (\ref{dm}) (see also (4.8) and (4.9) in~\cite{KP}).
The first term $(U^n\, \Sigma^{M *,{\rm UV}}\,  U^{n T})_{ll'}$ on the
RHS of  the last equality in~(\ref{proof1})  vanishes by itself,  as a
consequence  of  the   absence    of the operators     $\bar{\nu}_{lL}
\nu^C_{l'L}$ from the local Lagrangian. The second term vanishes, only
if
\begin{equation}
  \label{term2}
U^n\, \widehat{M}^n\, \Sigma^{L,{\rm UV}}\,U^{n T}\,\Big|_{ll'}\ =\ 0\,,
\end{equation}
or equivalently if 
\begin{equation}
  \label{term2new}
U^{n *}\, \Sigma^{L,{\rm UV}}\,U^{n T}\,\Big|_{\alpha l'}\ =\ 0\,,
\end{equation}
with  $\alpha   =  N_G   +   1,   N_G +    2,   \dots,   N_G +   N_R$.
Equation~(\ref{term2new}) is derived  by inserting the unity, $U^{n T}
U^{n   *}  =   {\bf   1}_{N_G+N_R}$,    between   $\widehat{M}^n$  and
$\Sigma^{L,{\rm UV}}$ in~(\ref{term2}), and noticing that $\big( U^n\,
\widehat{M}^n\, U^{n T}\big)_{li} = M^{n *}_{li} = m^*_{D\,l \alpha}$.
Employing the analytic expressions for  the neutrino self-energies  in
Appendix A (see also (\ref{MnuL}) below),  it is not difficult to show
that  (\ref{term2new}) is indeed  valid.   In fact,  the vanishing  of
$\big(U^{n *}\, \Sigma^{L,{\rm UV}}\,U^{n T}\big)_{\alpha l'}$ results
from  the  absence  of  the  lepton-number-violating   kinetic   terms
$\bar{\nu}_{lL} i \!  \!   \not  \!   \partial \nu^C_{\alpha R}$    of
dimension 4 from the original Lagrangian in the flavour space.  In the
SM with right-handed neutrinos,  the violation of lepton number occurs
softly through    the   Majorana  operators  $\bar{\nu}_{\alpha     R}
\nu^C_{\beta R}$  of dimension 3,  which is  reflected in the neutrino
mass-mixing     sum rule~(\ref{mixsum}).   This    completes our proof
of~(\ref{CTsum}).

We end  our discussion by  remarking that our  general approach to the
mixing-matrix    renormalization  may be   applied   to supersymmetric
theories as well,    e.g.\   to the  unitary   matrix~\cite{NPY}  that
diagonalizes the neutralino mass matrix  in the Minimal Supersymmetric
Standard Model (MSSM).    In this case,   a  convenient framework  for
renormalization that respects supersymmetry  is the so-called modified
dimensional    reduction    (${\overline {\rm DR}}$) scheme~\cite{DR}.
Alternatively, one  may work in  the ${\overline{\rm  MS}}$ scheme and
translate the  results into the ${\overline  {\rm DR}}$ scheme.  As in
the Majorana-neutrino case,   the  particular  zero  texture   in  the
neutralino  mass  matrix is protected  in the   ${\overline {\rm DR}}$
scheme, but needs be reinforced  by adding appropriate  anti-hermitian
constants to the ${\overline {\rm DR}}$-renormalized neutralino mixing
matrix, if the neutralino masses are renormalized in the OS scheme.

\setcounter{equation}{0}
\section{Renormalization-group equations}\label{sec:rg}

As an  immediate  application of  our  study in Sections~\ref{sec:mix}
and~\ref{sec:sum}, we derive  the gauge-independent RGEs of the mixing
matrices $B$ and $C$ in the SM  with right-handed neutrinos. With this
aim, we first  compute their respective  CTs $\delta B$ and $\delta C$
by means of~(\ref{dB})   and~(\ref{dC}) in the  ${\overline{\rm  MS}}$
scheme~\cite{KP1}:
\begin{eqnarray}
  \label{dBself}
\delta B^{\overline{\rm  MS}}_{li} &=& \sum\limits_{l'\neq l}^{N_G}
\frac{B_{l'i}}{2(m^2_l - m^2_{l'})}\, 
\Big[\, \big( m^2_l + m^2_{l'}\big)\, 
\Sigma^{l,L}_{ll'}\, +\, 2m_l m_{l'}\, \Sigma^{l,R}_{ll'} 
\, +\, 2\,\big(\, m_l \Sigma^{l,D}_{ll'} + m_{l'}
\Sigma^{l,D*}_{l'l}\,\big)\,\Big]^{\rm UV} \nonumber\\
&&\hspace{-1.8cm}-\! \sum\limits_{k\neq i}^{N_G+ N_R}\!\!
\frac{B_{lk}}{2(m^2_k - m^2_i)}\, \Big[\, \big( m^2_k + m^2_i \big)\, 
\Sigma^{n,L}_{ki}\, +\, 2m_k m_i\, \Sigma^{n,R}_{ki}\, +\, 
2\,\big(\, m_k \Sigma^{n,D}_{ki} + m_i
\Sigma^{n,D*}_{ik}\,\big)\,\Big]^{\rm UV}\!,\qquad\\
  \label{dCself}
\delta C^{\overline{\rm  MS}}_{ij} &=&\!\!\!\!\! 
\sum\limits_{k\neq i}^{N_G+N_R}\!
\frac{C_{kj}}{2(m^2_i - m^2_k)}\, 
\Big[\, \big( m^2_i + m^2_k\big)\, 
\Sigma^{n,L}_{ik}\, +\, 2m_i m_k\, \Sigma^{n,R}_{ik} 
\, +\, 2\,\big(\, m_i \Sigma^{n,D}_{ik} + m_k
\Sigma^{n,D*}_{ki}\,\big)\,\Big]^{\rm UV} \nonumber\\
&&\hspace{-1.8cm}-\! \sum\limits_{k\neq j}^{N_G+ N_R}\!\!
\frac{C_{ik}}{2(m^2_k - m^2_j)}\, \Big[\, \big( m^2_k + m^2_j \big)\, 
\Sigma^{n,L}_{kj}\, +\, 2m_k m_j\, \Sigma^{n,R}_{kj}\, +\, 
2\,\big(\, m_k \Sigma^{n,D}_{kj} + m_j
\Sigma^{n,D*}_{jk}\,\big)\,\Big]^{\rm UV}\! .
\end{eqnarray}
Note that the expressions (\ref{dBself}) and (\ref{dCself}) pertain to
Dirac neutrinos. In case  of Majorana neutrinos, these expressions are
supplemented  by  the  constraints stated  in~(\ref{Majconstr}),  with
$\Sigma^D_{ij}$ replaced by $\Sigma^M_{ij}$.

{}From  the   analytic  results  presented   in  Appendix  A,   it  is
straightforward to  deduce the  UV-divergent parts of  the self-energy
functions  occurring  in~(\ref{dBself})   and  (\ref{dCself})  in  the
$R_\xi$  gauge.   We  start  listing  the UV-divergent  parts  of  the
individual self-energy functions for the charged leptons
\begin{eqnarray}
  \label{lLdiv}
\big(\Sigma^{l,L}_{ll'}\big)^{\rm UV} &=& \frac{\alpha_w}{16\pi}\,
\bigg[\, \delta_{ll'}\,\bigg(\, 4s^2_w\xi_\gamma\: +\: 2\xi_W\: +\: 
\frac{(1-2s^2_w)^2}{c^2_w}\, \xi_Z\: +\: \frac{m^2_l}{M^2_W}\,
\bigg)\nonumber\\ 
&&+\, B_{li}B^*_{l'i}\, \frac{m^2_i}{M^2_W}\,\bigg]\, C_{\rm UV}\,,\\
  \label{LRdiv}
\big(\Sigma^{l,R}_{ll'}\big)^{\rm UV} &=& \frac{\alpha_w}{16\pi}\,
\delta_{ll'}\,\bigg(\, 4s^2_w\xi_\gamma\: +\: 
\frac{4s^4_w}{c^2_w}\, \xi_Z\: +\: 
\frac{2m^2_l}{M^2_W}\,\bigg)\,C_{\rm UV}\,,\\
  \label{lDdiv}
\big(\Sigma^{l,D}_{ll'}\big)^{\rm UV} &=& -\,\frac{\alpha_w}{16\pi}\, m_l\,
\bigg[\, \delta_{ll'}\,\bigg(\, 4s^2_w\,(3+\xi_\gamma)\: -\:
\frac{2s^2_w (1-2s^2_w)}{c^2_w}\, (3+\xi_Z)\: +\: \xi_W\: +\: 
\frac{\xi_Z}{2c^2_w}\, \bigg)\nonumber\\ 
&&+\, 2B_{li}B^*_{l'i}\, \frac{m^2_i}{M^2_W}\,\bigg]\, C_{\rm UV}\,.
\end{eqnarray}
Here  and in the following,  we consider the summation convention over
repeated indices in  their  whole  allowed range,   unless  explicitly
stated otherwise.  Specifically, charged  lepton indices, such  as $l$
and $l'$, are  summed  from 1 to $N_G$,  and  neutrino indices,  e.g.\
$i,j,k,n$,   from 1   to $N_G +N_R$.    The  divergent  pieces  of the
self-energy functions for Dirac neutrinos are given by
\begin{eqnarray}
  \label{DnuL}
\big(\Sigma^{n,L}_{ij}\big)^{\rm UV} &=& \frac{\alpha_w}{16\pi}\,
\bigg[\, C_{ij}\,\bigg(\, 2\xi_W\: +\: \frac{\xi_Z}{c^2_w}\,\bigg)\: +\: 
B^*_{li} B_{lj}\,\frac{m^2_l}{M^2_W}\: +\:
C_{ik}C_{kj}\,\frac{m^2_k}{M^2_W}\,\bigg]\, C_{\rm UV}\,,\\
  \label{DnuR}
\big(\Sigma^{n,R}_{ij}\big)^{\rm UV} &=& \frac{\alpha_w}{16\pi}\,
C_{ij}\, \frac{m_i m_j}{M^2_W}\, C_{\rm UV}\,,\\
  \label{DnuD}
\big(\Sigma^{n,D}_{ij}\big)^{\rm UV} &=& -\,\frac{\alpha_w}{16\pi}\, m_i\,
\bigg[\, C_{ij}\,\bigg(\,\xi_W\: +\: \frac{\xi_Z}{2c^2_w}\,\bigg)\: +\: 
2B^*_{li} B_{lj}\,\frac{m^2_l}{M^2_W}\,\bigg]\, C_{\rm UV}\,.
\end{eqnarray}
If  the  neutrinos  are  Majorana  particles, the  UV   parts  of  the
self-energy functions then read
\begin{eqnarray}
  \label{MnuL}
\big(\Sigma^{n,L}_{ij}\big)^{\rm UV} &=& \frac{\alpha_w}{16\pi}\,
\bigg[\, C_{ij}\,\bigg(\, 2\xi_W\: +\: \frac{\xi_Z}{c^2_w}\,\bigg)\: +\: 
B^*_{li} B_{lj}\,\frac{m^2_l}{M^2_W}\: +\: C^*_{ij}\, \frac{m_im_j}{M^2_W}
\nonumber\\ 
&&+\: C_{ik}C_{kj}\,\frac{m^2_k}{M^2_W}\,\bigg]\, C_{\rm UV}\,,\\
  \label{MnuM}
\big(\Sigma^{n,M}_{ij}\big)^{\rm UV} &=& -\,\frac{\alpha_w}{16\pi}\, 
\bigg[\, (m_iC_{ij}\,+\,m_jC^*_{ij})\,
\bigg(\,\xi_W\: +\: \frac{\xi_Z}{2c^2_w}\,\bigg)\nonumber\\
&& +\:  2(m_iB^*_{li} B_{lj} + m_j B_{li} B^*_{lj})\,
\frac{m^2_l}{M^2_W}\,\bigg]\, C_{\rm UV}\,.
\end{eqnarray}

The above  analytic results of the  UV pole structure  of the neutrino
self-energies reveal  that the RG  running of the mixing  matrices $B$
and $C$ depends on the nature of neutrinos, namely on whether they are
Dirac or  Majorana particles.  In  the ${\overline {\rm  MS}}$ scheme,
the  $\mu$-dependence of  $B$  and $C$  may  be computed  by the  beta
functions $\beta_B$ and $\beta_C$ as
\begin{equation}
  \label{BCRG}
\beta_B\ =\ \mu\, \frac{d B}{d\mu}\ =\ \lim\limits_{\varepsilon \to 0} 
\ \varepsilon g_w\, \frac{\partial }{\partial g_w}\,
\delta B^{\overline {\rm MS}}\,,\qquad
\beta_C\ =\ \mu\, \frac{d C}{d\mu}\ =\ \lim\limits_{\varepsilon \to 0} 
\ \varepsilon g_w\, \frac{\partial }{\partial g_w}\,
\delta C^{\overline {\rm MS}}\,,\quad
\end{equation}
where we  have employed the  fact that $\mu\, dg_w/d\mu = -\varepsilon
g_w  +  {\cal   O}  (g^3_w)$.    With  the  help  of~(\ref{BCRG})  and
of~(\ref{dBself}) and (\ref{dCself}), we obtain the following one-loop
beta functions for the Dirac neutrinos:
\begin{eqnarray}
  \label{betaBD}
\beta_{B_{li}} &=& \frac{\alpha_w}{16\pi}\, \bigg\{\,
\sum\limits^{N_G}_{l'\neq l}\,
\frac{m^2_l + m^2_{l'}}{m^2_{l'} - m^2_l}\, B_{lk}B^*_{l'k} B_{l'i}\, 
\frac{3m^2_k}{M^2_W}\nonumber\\
&&- \sum\limits_{k\neq i}^{N_G+N_R} \frac{B_{lk}}{m^2_k - m^2_i}\,
\bigg[\,(m^2_k + m^2_i)\, \bigg(\, C_{kn}C_{ni}\,\frac{m^2_n}{M^2_W}\:
-\: B^*_{lk}B_{li}\,\frac{3m^2_l}{M^2_W}\,\bigg)\nonumber\\
&& +\: 2C_{ki}\,\frac{m^2_k m^2_i}{M^2_W}\,\bigg]\,\bigg\}\,,\\
  \label{betaCD}
\beta_{C_{ij}} &=& \frac{\alpha_w}{16\pi}\, \bigg\{
\sum\limits_{k\neq i}^{N_G+N_R} \frac{C_{kj}}{m^2_i - m^2_k}\,
\bigg[\,(m^2_i + m^2_k)\, \bigg(\, C_{in}C_{nk}\,\frac{m^2_n}{M^2_W}\:
-\: B^*_{li}B_{lk}\,\frac{3m^2_l}{M^2_W}\,\bigg)\nonumber\\
&& +\: 2C_{ik}\,\frac{m^2_i m^2_k}{M^2_W}\,\bigg]\,\nonumber\\
&&- \sum\limits_{k\neq j}^{N_G+N_R} \frac{C_{ik}}{m^2_k - m^2_j}\,
\bigg[\,(m^2_k + m^2_j)\, \bigg(\, C_{kn}C_{nj}\,\frac{m^2_n}{M^2_W}\:
-\: B^*_{lk}B_{lj}\,\frac{3m^2_l}{M^2_W}\,\bigg)\nonumber\\
&& +\: 2C_{kj}\,\frac{m^2_k m^2_j}{M^2_W}\,\bigg]\,\bigg\}\,.
\end{eqnarray}
Correspondingly,    the   one-loop beta   functions   for the Majorana
neutrinos are given by
\begin{eqnarray}
  \label{betaBM}
\beta_{B_{li}} &=& \frac{\alpha_w}{16\pi}\, \bigg\{\,
\sum\limits^{N_G}_{l'\neq l}\,
\frac{m^2_l + m^2_{l'}}{m^2_{l'} - m^2_l}\, B_{lk}B^*_{l'k} B_{l'i}\, 
\frac{3m^2_k}{M^2_W}\nonumber\\
&&- \sum\limits_{k\neq i}^{N_G+N_R} \frac{B_{lk}}{m^2_k - m^2_i}\,
\bigg[\,(m^2_k + m^2_i)\, \bigg(\, C^*_{ki}\, \frac{m_k m_i}{M^2_W}\:
+\: C_{kn}C_{ni}\,\frac{m^2_n}{M^2_W}\:
-\: B^*_{lk}B_{li}\,\frac{3m^2_l}{M^2_W}\,\bigg)\nonumber\\
&& +\: 2m_km_i\,\bigg(\, C_{ki}\,\frac{m_k m_i}{M^2_W}
+\: C^*_{kn}C^*_{ni}\,\frac{m^2_n}{M^2_W}\:
-\: B_{lk}B^*_{li}\,\frac{3m^2_l}{M^2_W}\,\bigg)\,\bigg]\,\bigg\}\,,  \\
  \label{betaCM}
\beta_{C_{ij}} &=& \frac{\alpha_w}{16\pi}\, \bigg\{ \sum\limits_{k\neq
i}^{N_G+N_R} \frac{C_{kj}}{m^2_i - m^2_k}\,
\bigg[\,(m^2_i + m^2_k)\, \bigg(\, C^*_{ik}\, \frac{m_i m_k}{M^2_W}\:
+\: C_{in}C_{nk}\,\frac{m^2_n}{M^2_W}\:
-\: B^*_{li}B_{lk}\,\frac{3m^2_l}{M^2_W}\,\bigg)\nonumber\\
&& +\: 2m_im_k\,\bigg(\, C_{ik}\,\frac{m_i m_k}{M^2_W}
+\: C^*_{in}C^*_{nk}\,\frac{m^2_n}{M^2_W}\:
-\: B_{li}B^*_{lk}\,\frac{3m^2_l}{M^2_W}\,\bigg)\,\bigg]\nonumber\\
&&-\: \sum\limits_{k\neq j}^{N_G+N_R} \frac{C_{ik}}{m^2_k - m^2_j}\,
\bigg[\,(m^2_k + m^2_j)\, \bigg(\, C^*_{kj}\, \frac{m_k m_j}{M^2_W}\:
+\: C_{kn}C_{nj}\,\frac{m^2_n}{M^2_W}\:
-\: B^*_{lk}B_{lj}\,\frac{3m^2_l}{M^2_W}\,\bigg)\nonumber\\
&& +\: 2m_km_j\,\bigg(\, C_{kj}\,\frac{m_k m_j}{M^2_W}
+\: C^*_{kn}C^*_{nj}\,\frac{m^2_n}{M^2_W}\:
-\: B_{lk}B^*_{lj}\,\frac{3m^2_l}{M^2_W}\,\bigg)\,\bigg]\,\bigg\}\,.
\end{eqnarray}

It  is worth  commenting again  on the  fact that  the  beta functions
$\beta_B$ and $\beta_C$ in~(\ref{betaBD})--(\ref{betaCM}) become gauge
independent  in the  ${\overline  {\rm MS}}$  scheme,  only after  the
gauge-dependent tadpole terms proportional  to $M^2_H$ have been added
to the self-energy functions  $\Sigma^D$ (or $\Sigma^M$).  To the best
of our knowledge, the beta functions $\beta_B$ and $\beta_C$ represent
the most general results pertaining  to the one-loop RG-running of the
mixing matrices $B$ and $C$ in  the existing literature of the SM with
isosinglet neutrinos.  However, we should remark that the derived RGEs
for $\beta_B$  and $\beta_C$ are  only valid for energies  larger than
the  heaviest neutrino  mass.  We  have not  considered  the threshold
effects  due to the  decoupling~\cite{CP} of  the heavy  neutrinos, as
these effects highly depend  on the particular low-energy structure of
the  model~\cite{numodels,EL,othermodels,VS,CP} and will  therefore be
studied elsewhere.

\section{Conclusions}

We have  revisited the problem of gauge  dependence that occurs in the
renormalization of mixing matrices, within the  context of two generic
frameworks:   (i) the quark  sector of  the SM  and  (ii) the leptonic
sector of the SM with isosinglet neutrinos.  Although we confirmed the
earlier observations~\cite{GGM,KMS,BBS}  that an on-shell renormalized
mixing matrix   contains  gauge-dependent  terms,  we   have observed,
however, that these  terms are UV  finite  and have a  pure dispersive
form.  Because of  this  last fact,  we  have  found that these  naive
gauge-dependent terms can always be absorbed  into the definition of a
manifestly  gauge-invariant, but physically equivalent, mixing matrix,
where the latter  is evaluated within  a generalized  ${\overline {\rm
MS}}$    scheme  of  renormalization.   This    generalized scheme  of
renormalization is obtained by adding gauge-independent anti-hermitian
constants  to  a gauge-invariant, ${\overline  {\rm MS}}$-renormalized
mixing matrix (cf.\ (\ref{dVgen})  and (\ref{Ucn})).  In this way, the
different     renormalization        schemes   proposed       in   the
literature~\cite{DS,KP,GGM,KMS,BBS,YY,DK},    including   the       OS
scheme~\cite{DS,KP}, may be described  for appropriate choices of  the
anti-hermitian constants.

Our generalized ${\overline {\rm MS}}$ approach to the renormalization
of  mixing matrices   may also  be  conveniently  applied to  maintain
fundamental  global  or   local  symmetries   of   the  unrenormalized
Lagrangian.  For instance, our approach may be utilized to protect the
texture-zero      structure   of      the  Majorana-neutrino      mass
matrix~(\ref{Mnu}) or similar   constrained  structures of  predictive
neutrino-mass models.   Even  though such   additional symmetries  are
automatically preserved   in the ${\overline   {\rm MS}}$ scheme, they
become distorted by UV-finite terms and so  need be reinforced, if the
renormalized masses    are   evaluated within   other  renormalization
schemes, such as the  frequently adopted OS scheme.  Most importantly,
our approach   of   mixing-matrix renormalization may be    applied to
supersymmetric  theories  as  well.  In this  case, the  corresponding
generalized ${\overline {\rm DR}}$ approach may be used to renormalize
the   mixing  matrices  that  occur  in the   chargino and  neutralino
sectors~\cite{NPY},  as  well as   in  the  squark  and Higgs   scalar
sectors~\cite{APremark,GSH,YY} of the MSSM.

As a byproduct  of our study, we have  derived in Section~\ref{sec:rg}
the gauge-independent RGEs for the ${\overline {\rm MS}}$-renormalized
mixing matrices  in the SM with isosinglet  neutrinos.  The so-derived
RGEs  are valid  for energies  that are  higher than  the mass  of the
heaviest of  the heavy neutrinos. We  have not taken  into account the
decoupling  effects  due  to  heavy neutrino  thresholds,  since  they
crucially depend  on the low-energy  structure of the  model. However,
they  prove  important to  properly  describe  the  RG-running of  the
observed neutrino masses and  the neutrino-oscillation angles at lower
energies.  We  plan to  return to this  phenomenologically interesting
topic in the near future.


\newpage

\def\theequation{\Alph{section}.\arabic{equation}}
\begin{appendix}
\setcounter{equation}{0}
\section{Neutral and charged lepton self-energies}

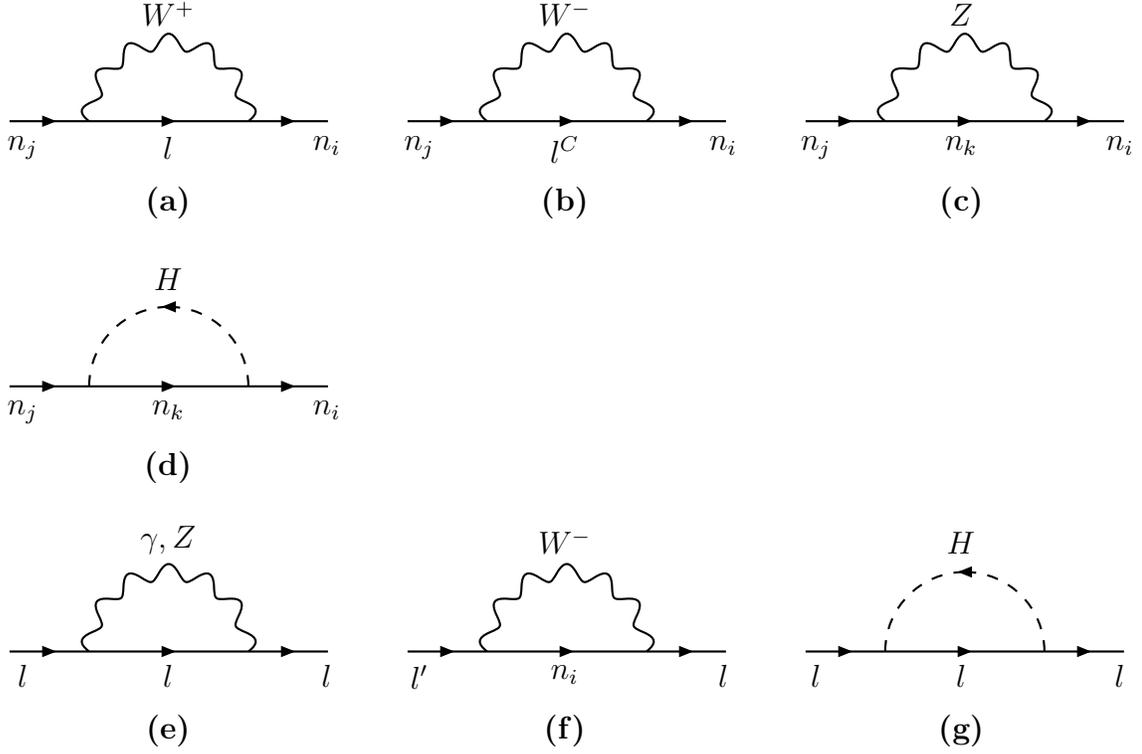
\begin{figure}[t]
\begin{center}
\begin{picture}(400,300)(0,0)
\SetWidth{0.8}

\ArrowLine(0,250)(30,250)\ArrowLine(30,250)(90,250)\ArrowLine(90,250)(120,250)
\PhotonArc(60,250)(30,0,180){3}{6.5}
\Text(5,245)[t]{$n_j$}\Text(60,245)[t]{$l$}\Text(120,245)[t]{$n_i$}
\Text(60,287)[b]{$W^+$}

\Text(60,220)[]{\bf (a)}

\ArrowLine(150,250)(180,250)\ArrowLine(180,250)(240,250)
\ArrowLine(240,250)(270,250)\PhotonArc(210,250)(30,0,180){3}{6.5}
\Text(155,245)[t]{$n_j$}\Text(210,245)[t]{$l^C$}\Text(270,245)[t]{$n_i$}
\Text(210,287)[b]{$W^-$}

\Text(210,220)[]{\bf (b)}

\ArrowLine(300,250)(330,250)\ArrowLine(330,250)(390,250)
\ArrowLine(390,250)(420,250)\PhotonArc(360,250)(30,0,180){3}{6.5}
\Text(305,245)[t]{$n_j$}\Text(360,245)[t]{$n_k$}\Text(420,245)[t]{$n_i$}
\Text(360,287)[b]{$Z$}

\Text(360,220)[]{\bf (c)}

\ArrowLine(0,150)(30,150)\ArrowLine(30,150)(90,150)\ArrowLine(90,150)(120,150)
\DashArrowArc(60,150)(30,0,180){4}
\Text(5,145)[t]{$n_j$}\Text(60,145)[t]{$n_k$}\Text(120,145)[t]{$n_i$}
\Text(60,187)[b]{$H$}

\Text(60,120)[]{\bf (d)}

\ArrowLine(0,50)(30,50)\ArrowLine(30,50)(90,50)\ArrowLine(90,50)(120,50)
\PhotonArc(60,50)(30,0,180){3}{6.5}
\Text(5,45)[t]{$l$}\Text(60,45)[t]{$l$}\Text(120,45)[t]{$l$}
\Text(60,87)[b]{$\gamma, Z$}

\Text(60,20)[]{\bf (e)}

\ArrowLine(150,50)(180,50)\ArrowLine(180,50)(240,50)
\ArrowLine(240,50)(270,50)\PhotonArc(210,50)(30,0,180){3}{6.5}
\Text(155,45)[t]{$l'$}\Text(210,45)[t]{$n_i$}\Text(270,45)[t]{$l$}
\Text(210,87)[b]{$W^-$}

\Text(210,20)[]{\bf (f)}

\ArrowLine(300,50)(330,50)\ArrowLine(330,50)(390,50)
\ArrowLine(390,50)(420,50)\DashArrowArc(360,50)(30,0,180){4}
\Text(305,45)[t]{$l$}\Text(360,45)[t]{$l$}\Text(420,45)[t]{$l$}
\Text(360,87)[b]{$H$}

\Text(360,20)[]{\bf (g)}

\end{picture}
\end{center}
\caption{\it Feynman graphs contributing to (a)--(d) neutral and
(e)--(f) charged lepton self-energies in the unitary gauge. If
neutrinos are Dirac particles, the graph (b) is absent.}\label{f1}
\end{figure}

Here,    we  present analytic    expressions  for   the  neutrino  and
charged-lepton self-energies in the renormalizable $R_\xi$ gauge.  The
Feynman diagrams that   contribute to the neutrino   self-energies are
shown in Fig.\ \ref{f1}(a)--(d),  while the corresponding graphs giving
rise   to  charged-lepton  self-energies   are   displayed    in Fig.\
\ref{f1}(e)--(g).  Our analytic results  are expressed in terms of the
usual Pasarino--Veltman one-loop functions \cite{PV}:
\begin{equation}
  \label{B0mu}
\{ B_0;\ B_\mu \,\}(p^2,m^2_1,m^2_2)\ =\ 
(2\pi\mu)^{4-D}\, \int\, \frac{d^Dk}{i\pi^2}\,
\frac{\{1;\ k_\mu\}}{(k^2 - m^2_1)\, [(k+p)^2 - m^2_2]}\ ,
\end{equation}
where the  Minkowski  space is  extended  to  $D =  4  - 2\varepsilon$
dimensions and $\mu$  is the so-called  't-Hooft mass scale.  Also, we
adopt the frequently-used 4-dimensional convention for the Minkowskian
metric  $g^{\mu \nu}$: $g^{\mu \nu}   = {\rm diag} (1,-1,-1,-1)$.  The
one-loop functions $B_0$  and  $B_\mu$, defined  in~(\ref{B0mu}),  are
given by
\begin{eqnarray}
  \label{B0} 
B_0 (p^2,m^2_1,m^2_2) &=& C_{\rm UV}\, -\, 
                     \ln \bigg(\frac{m_1m_2}{\mu^2}\bigg)\,
+\, 2\, +\, \frac{1}{p^2}\, \bigg[\, (m^2_2-m^2_1)\,
\ln\bigg(\frac{m_1}{m_2}\bigg)\nonumber\\ 
&&+\, \lambda^{1/2}(p^2,m^2_1,m^2_2)\,\, {\rm cosh}^{-1} \bigg(
\frac{m^2_1+m^2_2-p^2}{2m_1m_2}\bigg)\, \bigg]\, ,\\
  \label{Bmu}
B_\mu (p^2,m^2_1,m^2_2) &=& p_\mu\, B_1 (p^2,m^2_1,m^2_2)\, ,
\end{eqnarray}
with $C_{\rm UV} = \frac{1}{\varepsilon} - \gamma_E + \ln 4\pi$,
$\lambda (x,y,z) = (x -y -z)^2 - 4yz$ and
\begin{equation}
  \label{B1}
B_1 (p^2,m^2_1,m^2_2) \ =\ \frac{m^2_2 - m^2_1}{2p^2}\
\bigg( B_0 (p^2,m^2_1,m^2_2)\ -\ B_0 (0,m^2_1,m^2_2)\bigg)\ -\ 
\frac{1}{2}\, B_0 (p^2,m^2_1,m^2_2)\, .
\end{equation}
The one-loop function $B_0(p^2,m^2_1,m^2_2)$ evaluated at $p^2 = 0$
simplifies to 
\begin{equation}
  \label{B00}
B_0 (0, m^2_1, m^2_2)\ =\  C_{\rm UV}\:  -\: 
                     \ln \bigg(\frac{m_1m_2}{\mu^2}\bigg)\: +\: 1\: +\:
\frac{m^2_1 + m^2_2}{m^2_1 - m^2_2}\, \ln\bigg(\frac{m_2}{m_1}\bigg)\, .
\end{equation}
{}From this last expression, a useful identity relating the arguments
of the $B_0$-function at $p^2 = 0$ may easily be derived
\begin{equation}
  \label{B0id}
B_0 (0,m^2_1,m^2_2) \ =\ \frac{m^2_1}{m^2_1 - m^2_2}\, 
B_0 (0,m^2_1,m^2_1)\ -\  \frac{m^2_2}{m^2_1 - m^2_2}\, 
B_0 (0,m^2_2,m^2_2)\ +\ 1\, .
\end{equation}
Equation (\ref{B0id}) may be successfully employed  to check the gauge
independence of physical quantities.

We first derive  analytic expressions for the   case of Dirac  singlet
neutrinos. In the $R_\xi$ gauge, these are given by
\begin{eqnarray}
  \label{SnuL}
\Sigma^{n,L}_{ij} (p^2) \!\!&=&\!\! -\, \frac{\alpha_w}{8\pi} \bigg\{\
B^*_{li} B_{lj}\, \bigg[\, 2B_1 (p^2,m^2_l,M^2_W)\: +\:
B_0(p^2,m^2_l,M^2_W)\: +\: 1\nonumber\\
&&-\, \xi_W B_0 (p^2,m^2_l,\xi_W M^2_W)\ +\
\frac{p^2 - m^2_l}{M^2_W} \bigg( B_1 (p^2,M^2_W,m^2_l) - 
B_1 (p^2,\xi_W M^2_W,m^2_l) \bigg)\nonumber\\
&&+\, \frac{m^2_l}{M^2_W}\,
B_1(p^2,m^2_l,\xi_W M^2_W)\, \bigg]\nonumber\\
&&+\, \frac{1}{2c^2_w}\, C^*_{ki} C_{kj}\, 
\bigg[\, 2B_1 (p^2,m^2_k,M^2_Z)\: +\:
B_0(p^2,m^2_k,M^2_Z)\: +\: 1\nonumber\\
&&-\, \xi_Z B_0 (p^2,m^2_k,\xi_Z M^2_Z)\ +\ 
\frac{p^2 - m^2_k}{M^2_Z} \bigg( B_1 (p^2,M^2_Z,m^2_k)\ -\ 
B_1 (p^2,\xi_Z M^2_Z,m^2_k)\bigg)\nonumber\\
&&+\, \frac{m^2_k}{M^2_Z}\,
\bigg( B_1(p^2,m^2_k,\xi_Z M^2_Z)\: +\: B_1(p^2,m^2_k,M^2_H)\bigg)
\, \bigg]\, \bigg\}\, ,\\[0.4cm]
  \label{SnuR}
\Sigma^{n,R}_{ij} (p^2) \!\!&=&\!\! -\, \frac{\alpha_w}{8\pi} \bigg[\
B^*_{li} B_{lj}\, \frac{m_i m_j}{M^2_W}\, B_1 (p^2,m^2_l,\xi_W M^2_W) 
\nonumber\\ 
&&+\, \frac{1}{2c^2_w}\, C^*_{ki} C_{kj}\, \frac{m_i m_j}{M^2_Z}\, 
\bigg( B_1(p^2,m^2_k,\xi_Z M^2_Z)\: +\: B_1(p^2,m^2_k,M^2_H)\bigg)
\, \bigg]\, ,\\[0.4cm]
  \label{SnuD}
\Sigma^{n,D}_{ij} (p^2) \!\!&=&\!\! -\, \frac{\alpha_w}{8\pi}\, m_i\, 
\bigg\{\ B^*_{li} B_{lj}\, \frac{m^2_l}{M^2_W}\, B_0 (p^2,m^2_l,\xi_W M^2_W) 
\nonumber\\ 
&&+\, \frac{1}{2c^2_w}\, C^*_{ki} C_{kj}\, \frac{m_k^2}{M^2_Z}\, 
\bigg( B_0(p^2,m^2_k,\xi_Z M^2_Z)\: -\: B_0(p^2,m^2_k,M^2_H)\bigg)
\nonumber\\
&&+\,\frac{1}{2}\, C_{ij}\, \bigg[\, \xi_W\, 
\bigg( 1\: + \: B_0 (0,\xi_W M^2_W,\xi_W M^2_W)\bigg)\nonumber\\
&&+\, \frac{1}{2c^2_w}\, \xi_Z\, \bigg( 1\: +\: B_0 (0,\xi_Z M^2_Z,
\xi_Z M^2_Z)\bigg)\, \bigg]\, \bigg\}\,.
\end{eqnarray}
Note  that in (\ref{SnuD}) the  $p^2$-independent  terms represent the
$\xi$-dependent  part   of   the   tadpole    contributions.     These
contributions  are crucial, as  they restore the gauge independence in
the RG-running of neutrino masses and mixing angles.

Next,     we present    analytic   expressions  for  Majorana-neutrino
self-energies in the $R_\xi$ gauge:
\begin{eqnarray}
  \label{SMnuL}
\Sigma^{n,L}_{ij} (p^2) \!\!&=&\!\! -\, \frac{\alpha_w}{8\pi} \bigg\{\
B^*_{li} B_{lj}\, \bigg[\, 2B_1 (p^2,m^2_l,M^2_W)\: +\:
B_0(p^2,m^2_l,M^2_W)\: +\: 1\nonumber\\
&&-\, \xi_W B_0 (p^2,m^2_l,\xi_W M^2_W) \ +\
\frac{p^2 - m^2_l}{M^2_W} \bigg( B_1 (p^2,M^2_W,m^2_l) - 
B_1 (p^2,\xi_W M^2_W,m^2_l) \bigg)\nonumber\\
&&+\, \frac{m^2_l}{M^2_W}\,
B_1(p^2,m^2_l,\xi_W M^2_W)\, \bigg]\ +\ 
B_{li} B^*_{lj}\, \frac{m_i m_j}{M^2_W}\, B_1(p^2,m^2_l,\xi_W M^2_W) 
\nonumber\\
&&+\, \frac{1}{2c^2_w}\, C^*_{ki} C_{kj}\, 
\bigg[\, 2B_1 (p^2,m^2_k,M^2_Z)\: +\:
B_0(p^2,m^2_k,M^2_Z)\: +\: 1\nonumber\\
&&-\, \xi_Z B_0 (p^2,m^2_k,\xi_Z M^2_Z)\ +\ 
\frac{p^2 - m^2_k}{M^2_Z} \bigg( B_1 (p^2,M^2_Z,m^2_k)\ -\ 
B_1 (p^2,\xi_Z M^2_Z,m^2_k)\bigg)\,\bigg]\nonumber\\
&&+\, \frac{1}{2c^2_w}\, 
\frac{1}{M^2_Z}\,\Big( m_i C_{ki}\: +\: m_k C^*_{ki}\Big)
\Big( m_k C_{kj} + m_j C^*_{kj}\Big)\nonumber\\
&&\times\, \bigg( B_1(p^2,m^2_k,\xi_Z M^2_Z)\: +\: B_1(p^2,m^2_k,M^2_H)\bigg)
\, \bigg\}\, ,\\[0.4cm]
  \label{SMnuM}
\Sigma^{n,M}_{ij} (p^2) \!\!&=&\!\! -\, \frac{\alpha_w}{8\pi} \bigg\{\
 \frac{m^2_l}{M^2_W}\,\Big( m_i\, B^*_{li} B_{lj}\: +\:
m_j\,B_{li}B^*_{lj} \Big)\,B_0 (p^2,m^2_l,\xi_W M^2_W) \nonumber\\ 
&&-\, \frac{1}{2c^2_w}\, \bigg[\, m_k C_{ki}C_{kj}\, \bigg( 
3B_0 (p^2,m^2_k,M^2_Z)\: -\: 2\: +\: 
\xi_Z B_0(p^2,m^2_k,\xi_Z M^2_Z)\bigg)\nonumber\\
&&-\,\frac{m_k}{M^2_Z}\, \Big( m_i C^*_{ki} \, +\, m_k C_{ki}\Big)
\Big( m_k C_{kj}\, +\, m_j C^*_{kj} \Big)\nonumber\\
&&\times\, 
\bigg( B_0(p^2,m^2_k,\xi_Z M^2_Z)\: -\: B_0(p^2,m^2_k,M^2_H)\bigg)
\, \bigg]\,\nonumber\\
&&+\, \frac{1}{2}\, \Big( m_i C_{ij} \, +\, m_j C^*_{ij}\Big)
\,\bigg[\, \xi_W\, \bigg( 1\: +\:  
B_0 (0,\xi_W M^2_W,\xi_W M^2_W)\bigg)\nonumber\\
&&+\, \frac{1}{2c^2_w}\, \xi_Z\, \bigg( 1\: +\: B_0 (0,\xi_Z M^2_Z,
\xi_Z M^2_Z)\bigg)\, \bigg]\,  \bigg\}\, .
\end{eqnarray}
Likewise, we included in (\ref{SMnuM}) the $\xi$-dependent part of the
tadpole contributions.

Finally, for  completeness,  we   give the  charged-lepton  transition
amplitudes for $l' \to l$:
\begin{eqnarray}
  \label{SlL}
\Sigma^{l,L}_{ll'} (p^2) \!\!&=&\!\! -\, \frac{\alpha_w}{8\pi} \bigg\{\
2s^2_w \delta_{ll'}\, \bigg[\, 2B_1(p^2,m^2_l,\mu^2_\gamma)\: +\: 
B_0(p^2,m^2_l,\mu^2_\gamma)\: +\: 1\nonumber\\
&&-\, \xi_\gamma B_0 (p^2,m^2_l,\xi_\gamma \mu^2_\gamma)\ +\
\frac{p^2 - m^2_l}{\mu^2_\gamma} \bigg( B_1 (p^2,\mu^2_\gamma,m^2_l) - 
B_1 (p^2,\xi_\gamma \mu^2_\gamma,m^2_l) \bigg)\,\bigg]\nonumber\\
&&+\, \frac{(1-2s^2_w)^2}{2c^2_w}\, \delta_{ll'} 
\bigg[ 2B_1 (p^2,m^2_l,M^2_Z)\: +\:
B_0(p^2,m^2_l,M^2_Z)\: +\: 1\nonumber\\
&&-\, \xi_Z B_0 (p^2,m^2_l,\xi_Z M^2_Z)\ +\ 
\frac{p^2 - m^2_l}{M^2_Z} \bigg( B_1 (p^2,M^2_Z,m^2_l)\ -\ 
B_1 (p^2,\xi_Z M^2_Z,m^2_l)\bigg)\,\bigg]\nonumber\\
&&+\, \delta_{ll'}\, \frac{1}{2c^2_w}\, \frac{m^2_l}{M^2_Z}\,
\bigg( B_1(p^2,m^2_l,\xi_Z M^2_Z)\: +\: B_1(p^2,m^2_l,M^2_H)\bigg)
\nonumber\\
&&+\, B_{li} B^*_{l'i}\, \bigg[\, 2B_1 (p^2,m^2_i,M^2_W)\: +\:
B_0(p^2,m^2_i,M^2_W)\: +\: 1\nonumber\\
&&-\, \xi_W B_0 (p^2,m^2_i,\xi_W M^2_W)\ +\
\frac{p^2 - m^2_i}{M^2_W} \bigg( B_1 (p^2,M^2_W,m^2_i) - 
B_1 (p^2,\xi_W M^2_W,m^2_i) \bigg)\nonumber\\
&&+\, \frac{m^2_i}{M^2_W}\,
B_1(p^2,m^2_i,\xi_W M^2_W)\, \bigg]\, \bigg\}\, ,\\[0.4cm]
  \label{SlR}
\Sigma^{l,R}_{ll'} (p^2) \!\!&=&\!\! -\, \frac{\alpha_w}{8\pi}
\bigg\{\ 2s^2_w \delta_{ll'}\, \bigg[\, 2B_1(p^2,m^2_l,\mu^2_\gamma)\: +\: 
B_0(p^2,m^2_l,\mu^2_\gamma)\: +\: 1\nonumber\\
&&-\, \xi_\gamma B_0 (p^2,m^2_l,\xi_\gamma \mu^2_\gamma)\ +\
\frac{p^2 - m^2_l}{\mu^2_\gamma} \bigg( B_1 (p^2,\mu^2_\gamma,m^2_l) - 
B_1 (p^2,\xi_\gamma \mu^2_\gamma,m^2_l) \bigg)\, \bigg]\nonumber\\
&&+\, \frac{2s^4_w}{c^2_w}\, \delta_{ll'} 
\bigg[ 2B_1 (p^2,m^2_l,M^2_Z)\: +\:
B_0(p^2,m^2_l,M^2_Z)\: +\: 1\nonumber\\
&&-\, \xi_Z B_0 (p^2,m^2_l,\xi_Z M^2_Z)\ +\ 
\frac{p^2 - m^2_l}{M^2_Z} \bigg( B_1 (p^2,M^2_Z,m^2_l)\ -\ 
B_1 (p^2,\xi_Z M^2_Z,m^2_l)\bigg)\,\bigg]\nonumber\\
&&+\, \delta_{ll'}\, \frac{1}{2c^2_w}\, \frac{m^2_l}{M^2_Z}\,
\bigg( B_1(p^2,m^2_l,\xi_Z M^2_Z)\: +\: B_1(p^2,m^2_l,M^2_H)\bigg)
\nonumber\\
&&+\, B_{li} B^*_{l'i}\, \frac{m_l m_{l'}}{M^2_W}\, 
B_1 (p^2,m^2_i,\xi_W M^2_W)\, \bigg\}\, ,\\[0.4cm]
  \label{SlD}
\Sigma^{l,D}_{ll'} (p^2) \!\!&=&\!\! -\, \frac{\alpha_w}{8\pi}\, m_l\, 
\bigg\{\ 2s^2_w\,\delta_{ll'}\, \bigg( 3B_0(p^2,m^2_l,\mu^2_\gamma)\:
\: -\: 2\: +\: \xi_\gamma B_0(p^2,m^2_l,\xi_\gamma \mu^2_\gamma
)\bigg)\nonumber\\  
&&-\, \frac{s^2_w(1-2s^2_w)}{c^2_w}\, \delta_{ll'} \bigg(
3B_0(p^2,m^2_l,M^2_Z)\: -\: 2\: +\: \xi_Z
B_0(p^2,m^2_l,\xi_Z M^2_Z )\bigg)\nonumber\\
&&+\, \delta_{ll'}\, \frac{1}{2c^2_w}\, \frac{m_l^2}{M^2_Z}\, 
\bigg( B_0(p^2,m^2_l,\xi_Z M^2_Z)\: -\:
B_0(p^2,m^2_l,M^2_H)\bigg)\nonumber\\
&& +\, 
B_{li} B^*_{l'i}\, \frac{m^2_i}{M^2_W}\, B_0 (p^2,m^2_i,\xi_W M^2_W)\
+\ \frac{1}{2}\, \delta_{ll'}\, \bigg[\, \xi_W\, 
\bigg( 1\: + \: B_0 (0,\xi_W M^2_W,\xi_W M^2_W)\bigg)\nonumber\\
&&+\, \frac{1}{2c^2_w}\, \xi_Z\, \bigg( 1\: +\: B_0 (0,\xi_Z M^2_Z,
\xi_Z M^2_Z)\bigg)\, \bigg]\, \bigg\}\,.
\end{eqnarray}
Apart from the $\xi$-dependent part of the  Higgs tadpoles included in
(\ref{SlD}), (\ref{SlL})--(\ref{SlD}) agree  well with those presented
in \cite{KMS}.

\end{appendix}


\begin{thebibliography}{99}
\bibitem{KP1} In the OS scheme, these expressions were derived in~\cite{KP}.
\end{thebibliography}
\end{document}